\documentclass[12pt]{article}

 \usepackage{amssymb}
    \usepackage{latexsym}

\newcommand{\bc}{\begin{center}}
\newcommand{\ec}{\end{center}}
\def\ba#1{\begin{array}{#1}\displaystyle}
\newcommand{\ea}{\end{array}}
\newcommand{\z}{\\[2mm] \displaystyle}
\newcommand{\ds}{\displaystyle}
\newcommand{\beq}{\begin{equation}}
\newcommand{\eeq}{\end{equation}}
\newcommand{\beqa}{\begin{eqnarray}}
\newcommand{\eeqa}{\end{eqnarray}}
\newcommand{\no}{\nonumber}
\newcommand{\n}{\nonumber\\}
\newcommand{\bi}{\begin{itemize}}
\newcommand{\ei}{\end{itemize}}

\def\sect#1{\section{#1}\setcounter{equation}{0}}
\def\ssect#1{\subsection{#1}}

\def\lt#1{\left#1}
\def\rt#1{\right#1}
\def\t#1{\tilde{#1}}
\def\h#1{\hat{#1}}
\def\b#1{\bar{#1}}
\def\frc#1#2{\frac{#1}{#2}}

\newcommand{\p}{\partial}
\newcommand{\prin}{\underline{\mathrm{P}}}

\newcommand{\vac}{{\rm vac}}
\newcommand{\bra}{\langle}
\newcommand{\ket}{\rangle}
\newcommand{\Z}{{\mathbb{Z}}}

\newcommand{\Or}{{\cal O}}

\newcommand{\ep}{\epsilon}

\newcommand{\Tr}{{\rm Tr}}

\newcommand{\Res}{{\rm Res}}
\newcommand{\End}{{\rm End}}

\newcommand{\ft}{{\cal L}}
\newcommand{\braL}{\langle\langle}
\newcommand{\ketL}{\rangle\rangle_L}

\begin{document}

\pagestyle{empty} \pagenumbering{arabic}

\vspace{3cm} \bc \Large\bf Finite-temperature form factors\\ in
the free Majorana theory \ec \vspace{2cm} \bc {\large Benjamin
Doyon}

{\em Rudolf Peierls Centre for Theoretical Physics\\
University of Oxford\\
1 Keble Road, Oxford OX1 3NP, U.K.} \ec \vspace{2cm} \bc \bf
Abstract \ec We study the large distance expansion of correlation
functions in the free massive Majorana theory at finite
temperature, alias the Ising field theory at zero magnetic field
on a cylinder. We develop a method that mimics the spectral
decomposition, or form factor expansion, of zero-temperature
correlation functions, introducing the concept of
``finite-temperature form factors''. Our techniques are different
from those of previous attempts in this subject. We show that an
appropriate analytical continuation of finite-temperature form
factors gives form factors in the quantization scheme on the
circle. We show that finite-temperature form factor expansions are
able to reproduce expansions in form factors on the circle. We
calculate finite-temperature form factors of non-interacting
fields (fields that are local with respect to the fundamental
fermion field). We observe that they are given by a mixing of
their zero-temperature form factors and of those of other fields
of lower scaling dimension. We then calculate finite-temperature
form factors of order and disorder fields. For this purpose, we
derive the Riemann-Hilbert problem that completely specifies the
set of finite-temperature form factors of general twist fields
(order and disorder fields and their descendants). This
Riemann-Hilbert problem is different from the zero-temperature
one, and so are its solutions. Our results agree with the known
form factors on the circle of order and disorder fields.
\vfill\noindent September 2005

\newpage
\pagestyle{plain} \setcounter{page}{1}

\sect{Introduction}

Quantum field theory (QFT) at finite temperature is a subject of
great interest which has been studied from many viewpoints (see,
for instance, \cite{Kapusta}). The main goal of QFT is the
reconstruction of a set of correlation functions in which the
physical information provided by the theory is embedded. For
instance, two-point correlation functions are related to response
functions, which can be measured and which provide precise
information about the dynamics of a physical system at
thermodynamic equilibrium. For the purpose of comparison with
experiment, it is important to know the influence of a non-zero
temperature on correlation functions. In particular, both static
(equal-time) and dynamical two-point correlation functions at
finite temperature still need more accurate study.

In recent years, thanks to advances in experimental techniques
allowing the identification and study of quasi-one-dimensional
systems (see for instance \cite{BourbonnaisJ99,Gruner}), there has
been an increased interest in calculating correlation functions in
1+1-dimensional integrable models of QFT (for applications of
integrable models to condensed matter systems, see for instance
the recent review \cite{EsslerK04}). Integrable models are of
particular interest, because in many cases, matrix elements of
local fields in eigenstates of the Hamiltonian, or form factors,
can be evaluated exactly by solving an appropriate Riemann-Hilbert
problem in the rapidity space
\cite{VergelesG76,Weisz77,KarowskiW78,BergKW79,Smirnov}. At zero
temperature, correlation functions are vacuum expectation values
in the Hilbert space of quantization on the line. A useful
representation of zero-temperature correlation functions is then
provided by their form factor expansion (or spectral
decomposition):
\[\ba{rl}
    \bra\vac|\Or(x,\tau)\Or(0,0)|\vac\ket =& \ds
    \sum_{k=0}^\infty\sum_{\ep_1,\ldots,\ep_k}
    \int\frc{d\theta_1\cdots d\theta_k}{k!}e^{-r\sum_jM(\ep_j)\cosh(\theta_j)}
    \times
    \z & \qquad\times
    \bra\vac|\Or(0,0)|\theta_1,\ldots,\theta_k\ket^{in}_{\ep_1,\ldots,\ep_k}
    {\ }^{\hspace{6mm}in}_{\ep_1,\ldots,\ep_k}
    \bra\theta_1,\ldots,\theta_k|\Or(0,0)|\vac\ket \ea
\]
where $\tau$ is the Euclidean time and $r=\sqrt{x^2+\tau^2}$. This
expansion is obtained by inserting the resolution of the identity
in terms of a basis of common eigenstates of the momentum operator
and of the Hamiltonian, parameterized by the rapidity variables
$\theta_j$'s and by the particle types $\ep_j$'s. Here $M(\ep)$ is
the mass of a particle of type $\ep$ (and we took the basis of {\em in}-states).
Using many of such insertions, form factor expansions can be
obtained for multi-point correlation functions as well. This is a
useful representation because it is a large-distance expansion,
which is hardly accessible by perturbation theory, and which is
often the region of interest in condensed matter applications.
Also, form factor expansions in integrable models at zero
temperature have proven to provide a good numerical accuracy for
evaluating correlation functions in a wide range of energies, and
combined to conformal perturbation theory give correlation
functions at all energy scales (an early work on this is
\cite{ZamolodchikovAl91}).

In this paper we study the large-distance expansion of correlation
functions in the free massive Majorana field theory on a line at
finite temperature. This model of 1+1-dimensional quantum field
theory occurs, for instance, as the off-critical scaling limit of
the finite-temperature Ising quantum chain (with an infinite
number of sites) in a transverse magnetic field, near to its
quantum critical point. It also occurs as the off-critical scaling
limit of the statistical Ising model on a cylinder, near to its
thermal critical point; the circumference of the cylinder is the
inverse temperature of the Majorana theory. For an exposition on
these scaling limits and for references of early works, see, for
instance, the book \cite{ItzyksonDrouffe}. In particular, in both
of these examples, the fields of most interest are the twist
fields associated to the $\Z_2$ symmetry of the Majorana theory
\cite{KadanofC71,SchroerT78}. These fields are the scaling limit
of the spin operator in the Ising chain (the Pauli matrix that
enjoys a nearest-neighbor interaction) and of the spin variables
in the statistical Ising model. There are two of these twist
fields: the order field and the disorder field. They can be seen
as representing the scaling limit of the same operator, but in
different regimes (ordered and disordered) away from the critical
point. When the mass is set to zero (at criticality), the theory
reduces to the well-known Ising conformal field theory.

The correlation functions of twist fields exhibit non-trivial,
``non-free'' behaviors, and can be obtained as appropriate
solutions to non-linear differential equations \cite{WuMTB76}. At
finite temperature, or on the cylinder, these equations are
partial differential equations in the coordinates on the cylinder
\cite{Perk80,Lisovyy02,FonsecaZ03}, and do not offer a very useful
tool for numerically evaluating correlation functions, neither for
analyzing their large-distance behavior (short-distance behaviors,
on the other hand, can always be analyzed by conformal
perturbation theory \cite{ZamolodchikovAl91}). Hence it is worth
studying in more detail the large-distance expansion of
correlation functions of twist fields. This study constitutes a
first step towards generalizations to interacting integrable
models. In particular, we will present a method that parallels the
zero-temperature form factor expansion.

\ssect{Correlation functions at finite temperature}

At finite temperature, correlation functions are obtained by
taking a statistical average of quantum averages, with Boltzmann
weights $e^{-LE}$ where $E$ is the energy of the quantum state:
\beq\label{ftcorr}
    \braL \Or(x,\tau)\cdots\ketL = \frc{\Tr\lt[e^{-LH}
    \Or(x,\tau)\rt]}{\Tr\lt[e^{-LH}\rt]}~.
\eeq
Here, $L$ is the inverse temperature and $H$ is the Hamiltonian. A
consequence of the imaginary-time formalism \cite{Matsubara55} is
the Kubo-Martin-Schwinger (KMS) identity \cite{Kubo57,MartinS59},
\beq\label{KMS}
    \braL \Or(x,\tau) \cdots \ketL =
    (-1)^f \braL \Or(x,\tau+L) \cdots \ketL
\eeq
where $f$ is the statistics of $\Or$ (it is 1 for fermionic
operators and 0 for bosinic operators), and where the dots
($\cdots$) represent local fields at time $\tau$ and at positions
different from $x$. Then, finite-temperature correlation functions
can be seen as vacuum expectation values in the same model of QFT,
but this time quantized on a circle of circumference $L$:
\beq\label{cylcorr}
    \braL \Or(x,\tau) \cdots \ketL = e^{i\pi s/2}\,{\ }_L\bra\vac| \Or_L(-\tau,x) \cdots
    |\vac\ket_L
\eeq
where $s$ is the spin of $\Or$. The operator $\Or_L(-\tau,x)$ is
the corresponding operator acting on the Hilbert space on the
circle, with space variable $-\tau$ (parameterizing the circle of
circumference $L$) and Euclidean time variable $x$ (going along
the cylinder). The vector $|\vac\ket_L$ is the vacuum in the
Hilbert space on the circle. There are two sectors in the
quantization on the circle: Neveu-Schwartz (NS) and Ramond (R),
where the fermion fields are anti-periodic and periodic,
respectively, around the circle. The trace (\ref{ftcorr}) with
insertion of operators that are local with respect to the fermion
fields naturally corresponds to the NS sector due to the KMS
identity. With insertion of twist fields, however, the sector may
be changed to R or to a mixed sector by an appropriate choice of
the associated branch cuts. The phase factor in (\ref{cylcorr}) is
in agreement with the fact that the operator $\Or_L$ is Hermitian
on the Hilbert space on the circle, if the corresponding operator
$\Or$ is Hermitian on the Hilbert space on the line.

\ssect{Previous works}

A straightforward application of the zero-temperature ideas would
suggest a form factor expansion of the vacuum expectation values
on the circle (\ref{cylcorr}):
\beq\label{cylffexp}\ba{l}
    {\ }_L\bra\vac|\Or_L(x,\tau)\Or_L(0,0)|\vac\ket_L = \z \qquad\quad \sum_{k=0}^\infty
    \sum_{n_1,\ldots,n_k} \frc{e^{\sum_j n_j \frc{2\pi i x}L - E_{n_1,\ldots,n_k}
    \tau}}{k!}
    {\ }_L\bra\vac|\Or_L(0,0)|n_1,\ldots,n_k\ket_L
    {\ }_L\bra n_1,\ldots,n_k|\Or_L(0,0)|\vac\ket_L \ea
\eeq
where the eigenstates of the momentum operator and of the
Hamiltonian on the circle are parametrized by discrete variables
$n_j$'s. This form is valid for any integrable models on the
circle; it can also obviously be generalized to other multi-point
correlation functions. In general integrable models, the energy
levels $E_{n_1,\ldots,n_k}$ are not known in closed form (although
exact methods exist to obtain non-linear integral equations that
define them: from thermodynamic Bethe ansatz techniques, from
calculations \`a la Destri-de Vega and from the so-called BLZ
program). Also the matrix elements of local fields ${\
}_L\bra\vac|\Or_L(0,0)|n_1,\ldots,n_k\ket$ (form factors on the circle)
still seem to be far from reach (however, see
\cite{Smirnov98a,Smirnov98b,ElburgS00,MussardoRS03}). Hence, this
method does not seem to be applicable in general yet.
Nevertheless, in the Majorana fermion theory with mass $m$, the
energy levels are simply given by
\[
    E_{n_1,\ldots,n_k} = \sum_{j=1}^k \sqrt{m^2+\lt(\frc{2\pi n_j}L\rt)^2}
\]
where $n_j\in\Z+\frc12$ in the Neveu-Schwartz (NS) sector and
$n_j\in\Z$ in the Ramond (R) sector (the sectors where the fermion
field have anti-periodic and periodic conditions around the
cylinder, respectively). Matrix elements of the primary order and
disorder fields were calculated in \cite{Bugrij00},
\cite{Bugrij01} in the lattice Ising model, and in a simpler way
in \cite{FonsecaZ01} directly in the Majorana field theory using
the free-fermion equations of motion and the ``doubling trick''.
The form factor expansion on the circle can then be used to obtain
the leading exponentially decreasing behavior of static
correlation functions in the quantum Ising chain, and, with
slightly more work, to obtain the large-time dynamical correlation
functions in the semi-classical regime \cite{AltshulerT05}. This
reproduces early results of \cite{Sachdev96} and
\cite{SachdevY97}.

A second set of ideas was suggested by Leclair, Lesage, Sachdev
and Saleur in \cite{LeclairLSS96}, starting from the
representation (\ref{ftcorr}). By partly performing the trace and
by using general properties of matrix elements of local fields,
the following representation for two-point functions in the
Majorana theory of mass $m$ was obtained:
\beq\label{ftffexpLLSS}\ba{l}
    \braL\Or_1(x,\tau)\Or_2^\dag(0,0)\ketL =\z
        \sum_{k=0}^\infty \sum_{\ep_1,\ldots,\ep_k=\pm}
    \int \frc{d\theta_1\cdots
    d\theta_k\;e^{\sum_{j=1}^k\ep_j (imx\sinh\theta_j-m\tau\cosh\theta_j)}}{
    k!\;\prod_{j=1}^k\lt(1+e^{-\ep_j
    mL\cosh\theta_j}\rt)}\;
    F^{\Or_1}_{\ep_1,\ldots,\ep_k}(\theta_1,\ldots,\theta_k)
    (F^{\Or_2}_{\ep_1,\ldots,\ep_k}(\theta_1,\ldots,\theta_k))^*
    \ea
\eeq
where $F^{\Or}_{\ep_1,\ldots,\ep_k}(\theta_1,\ldots,\theta_k)$ are
matrix elements of $\Or$ in the Hilbert space on the line of the
form
$\bra\vac|\Or|\theta_{j_1},\ldots,\theta_{j_n},\theta_{i_m}+i\pi,\ldots,\theta_{i_1}+i\pi\ket$,
where all $\theta_j$'s correspond to $\ep_j=(+)$, and all
$\theta_i$'s correspond to $\ep_i=(-)$.

This expression is better adapted to study dynamical correlation
functions in real time $\tau=it$ than the form factor expansion on
the circle. Another advantage of this idea is that, following
methods similar to those used in \cite{BabelonB92} at zero
temperature, one can obtain a Fredholm determinant representation
for finite-temperature two-point functions, which can be useful
for further analysis. Such a representation was obtained in
\cite{LeclairLSS96} and was used in order to obtain a set of
non-linear partial differential equations for the two-point
function of twist fields. This lead to the known equations plus,
in particular, an additional equation involving temperature
derivatives. However, this additional equation does not follow
from any of the other methods used in other works
\cite{Perk80,Lisovyy02,FonsecaZ03}, and it was suggested not to
hold in \cite{FonsecaZ03}.

This idea also seems to allow more easily generalizations to
interacting integrable models. Such a generalization was
conjectured by Leclair and Mussardo in \cite{LeclairM99}. Saleur,
however, pointed out that this generalization might not be correct
\cite{Saleur00}, and the numerical calculation of Castro-Alvaredo
and Fring in the scaling Lee-Yang model tended to agree with this
incorrectness \cite{Castro-AlvaredoF02}.

Note that verifications of the results of \cite{LeclairLSS96} were
made for some fields in the complex (Dirac) free fermion and in
the Federbush model in \cite{Castro-AlvaredoF02}, and that
agreement was found with known results in various limits. Leclair
and Mussardo also proposed in \cite{LeclairM99}, along similar
lines, a formula for one-point functions of local operators in
interacting integrable models, which, after some controversy, seem
to be confirmed \cite{Delfino01,Mussardo01,Lukyanov01}, although
it is fair to say that more checks would be appropriate (on the
subject of one-point functions from form factors, see also
\cite{Balog94}).

In the works mentioned above, very little was said concerning
finite-temperature correlation functions with more than two
operators.

\ssect{Our work}

In the following, we will develop a scheme for obtaining
large-distance representations of finite-temperature correlation
functions in the Majorana theory of the form:
\beq\label{ftffexp}\ba{l}
    \braL\Or_1(x,\tau)\Or_2^\dag(0,0)\ketL =\z
        \sum_{k=0}^\infty \sum_{\ep_1,\ldots,\ep_k=\pm}
    \int \frc{d\theta_1\cdots
    d\theta_k\;e^{\sum_{j=1}^k\ep_j (imx\sinh\theta_j-m\tau\cosh\theta_j)}}{
    k!\;\prod_{j=1}^k\lt(1+e^{-\ep_j
    mL\cosh\theta_j}\rt)}\;
    f^{\Or_1}_{\ep_1,\ldots,\ep_k}(\theta_1,\ldots,\theta_k;L)
    (f^{\Or_2}_{\ep_1,\ldots,\ep_k}(\theta_1,\ldots,\theta_k;L))^*
    \ea
\eeq
with appropriate generically $L$-dependent functions
$f^\Or_{\ep_1,\ldots,\ep_k}(\theta_1,\ldots,\theta_k;L)$. We will
see that this formula is valid:
\begin{itemize}
\item with both $\Or_1,\,\Or_2$ fields that are local with respect to
the fermion fields, in the NS sector;
\item with one of $\Or_1$ or $\Or_2$ a twist field, in a mixed R-NS or NS-R sector;
\item with both $\Or_1,\,\Or_2$ twist fields, in the R sector.
\end{itemize}
This scheme does not say anything new about one-point functions;
in particular, the functions $f^\Or$ are evaluated up to
normalization using analytic properties in rapidity space. We will
only analyze two-point correlation functions, but the techniques
and ideas introduced can be easily generalized to multi-point
correlation functions.

Although our expansion (\ref{ftffexp}) for two-point functions is
of the same form as the expansion (\ref{ftffexpLLSS}), we will
find that our methods for obtaining (\ref{ftffexp}) is clearly
more powerful than an explicit evaluation of the trace as in
\cite{LeclairLSS96}, and more importantly that most of the results
of \cite{LeclairLSS96} (where this form first appeared) are in fact incorrect\footnote{Note that in the
latter work, some singularities were neglected
without good reasons (as was admitted there) in the derivation of (\ref{ftffexpLLSS}).}.
It is a simple matter to check that the positive verifications of
the results of \cite{LeclairLSS96} made in the works described in
the previous subsection, being only for simple fields or in
particular limits, can still be made on the formulas that we
propose. We will not do that explicitly here, since we have a much
stronger verification of the correctness of our results through
their relation with the uncontroversial form factors on the
circle. Our main results can be grouped into 4 points that exhibit
the differences between our approach and results, and those of
previous works. They are:

{\bf 1. Definition of finite-temperature form factors.} The
objects $f^{\Or}_{\ep_1,\ldots,\ep_k}(\theta_1,\ldots,\theta_k;L)$
involved in (\ref{ftffexp}), which we will call
``finite-temperature form factors'', are {\em not} given by simple
matrix elements $F^\Or$ of $\Or$ in the Hilbert space on the line
as was suggested in the proposal \cite{LeclairLSS96}. They are
rather given by appropriate traces of multiple commutators
($[\cdot,\cdot]$) and anti-commutators ($\{\cdot,\cdot\}$) on the
Hilbert space of the Majorana theory on the line:
\beq\label{ftffcac}
    f^{\Or}_{\ep_1,\ldots,\ep_k}(\theta_1,\ldots,\theta_k;L)
    =
    \braL
    \{a^{\ep_1}(\theta_1),[a^{\ep_2}(\theta_2),\{\cdots,\Or(0,0)\cdots\}]\}\ketL\quad
    (\theta_i\neq\theta_j\;\forall\;i\neq j)
\eeq
where $a^+(\theta) \equiv a^\dag(\theta)$ is a creation operator
for a particle of rapidity $\theta$, and $a^-(\theta) \equiv
a(\theta)$ is an annihilation operator for a particle of rapidity
$\theta$. Simple calculations with normal-ordered products of
fermion fields, for instance, make it clear that generically,
these objects are not equal to zero-temperature form factors. The
difference will be explained in Point 3 below.

The techniques we use to derive such a representation are related
to the physical theory of particle and hole excitations above a
``thermal vacuum'', which was initially proposed more than thirty
years ago and which developed into a mature theory under the name
of thermo-field dynamics \cite{LeplaeUM74,ArimitsuU87} (for a
review, see for instance \cite{Henning95}). We will explain how
the traces (\ref{ftffcac}) can be seen as matrix elements between
a thermal vacuum (on the left) and a state with particle and hole
excitations (on the right), and how the representation
(\ref{ftffexp}) comes from a resolution of the identity on the
space of NS-sector thermal states. In fact, there are still very
few applications of the ideas of integrable quantum field theory
(like the form factor program) to thermo-field dynamics (see,
however, the recent study \cite{AmaralB05} of bosonization in
thermo-field dynamics). Our study constitutes a step in this
direction.

{\bf 2. Relation between form factors on the circle and
finite-temperature form factors.} We will show that an appropriate
analytical continuation in rapidity space of the
finite-temperature form factors reproduces form factors on the
circle. This links the two sets of ideas mentioned above, and, as
we will see below, gives an ``analytical'' way of calculating form
factors on the circle. The precise correspondence is given by
\beq\label{relftcyl}\ba{l}\ds
    {\ }_L\bra\t{n}_1,\ldots,\t{n}_l|\Or_L(0)|n_1,\ldots,n_k\ket_L = e^{-\frc{i\pi
    s}2}
        \lt(\frc{2\pi}{m L}\rt)^{\frc{k+l}2}
        \lt(\prod_{j=1}^l \frc1{\sqrt{\cosh(\alpha_{\t{n}_j})}}\rt)\;
        \lt(\prod_{j=1}^k
        \frc1{\sqrt{\cosh(\alpha_{n_j})}}\rt)\;\times\z \qquad\qquad\qquad\qquad\times\;
        f^\Or_{+,\ldots,+,-,\ldots,-}\lt(\alpha_{n_1}+\frc{i\pi}2,\ldots,
        \alpha_{n_k}+\frc{i\pi}2,
        \alpha_{\t{n}_l}+\frc{i\pi}2,\ldots,\alpha_{\t{n}_1}+\frc{i\pi}2;L\rt)
    \ea
\eeq
where there are $k$ positive charges and $l$ negative charges in
the indices of $f^\Or$, and where $\alpha_n$ are defined by
\beq\label{defalphan}
    \sinh\alpha_n = \frc{2\pi n}{m L} \quad (n\in\Z+\frc12)~.
\eeq
This formula is valid for any excited states in the NS sector.
When $\Or$ is a twist field, formula (\ref{relftcyl}) can then be
applied only if one of the bra or the ket is the vacuum, and if
the branch associated to this twist field is chosen so that this
vacuum is in the R sector and the excited state is in the NS
sector. Using this formula, we will show that the form factor
expansion on the circle and the finite-temperature form factor
expansion give the same result.

{\bf 3. Evaluation of finite-temperature form factors.} We will
evaluate explicitly the finite-temperature form factors
$f^{\Or}_{\ep_1,\ldots,\ep_k}(\theta_1,\ldots,\theta_k;L)$ for
some fields. We find:
\begin{enumerate}
\item {\em Mixing.} For any local non-interacting field $\Or$ (a field that is
local with respect to the fundamental fermion field), they are
equal to matrix elements in the Hilbert space on the line of a sum
of a finite number of local fields, including $\Or$ as well as
operators of lower scaling dimensions and of equal or lower
conformal dimensions. We will present a procedure for calculating
the associated mixing matrix. This mixing can ultimately be seen
as coming from the difference between the normal-ordering on the
line and the normal-ordering on the circle. Note that in the
proposal \cite{LeclairLSS96}, only the energy field, among
non-interacting fields, was studied, and the mixing was treated by
adding to this field the identity operator in such a way that it
have zero thermal expectation value. For other fields, such a
simple prescription would not be enough.
\item {\em Leg factors.} For the order and disorder fields, they are
given by their matrix elements in the Hilbert space on the line
times functions depending on the individual rapidities (``leg
factors''). These factors were absent in \cite{LeclairLSS96}.
\end{enumerate}
Note that, consistently with (\ref{relftcyl}), both the mixing
phenomenon and the appearance of leg factors are observed in form
factors on the circle. In particular, we were able to reproduce
from our results the known form factors on the circle.

We evaluate the finite-temperature form factors of order and
disorder fields by deriving and solving a Riemann-Hilbert problem
that completely fixes the set of finite-temperature form factors
of all local twist fields (order and disorder fields and their
descendants) in the Majorana field theory. The Riemann-Hilbert
problem depends explicitly on the temperature. For descendant of
order and disorder fields, we present a conjecture for the
structure in which the mixing phenomenon can be described
(without, though, calculating the associated mixing matrix).

It is important to note that since the effect of temperature on
the finite-temperature form factor of order and disorder fields is
only in the leg factors, a Fredholm determinant representation can
straightforwardly be obtained from the finite-temperature form
factor expansion, as it was done in \cite{LeclairLSS96}. One needs
only replace the filling fraction of \cite{LeclairLSS96} by the
filling fraction times our leg factors. However, because of these
factors, following the derivation of \cite{LeclairLSS96} it is
clear that the equation involving temperature derivatives is
incorrect, in accordance with the suggestion of \cite{FonsecaZ03}.

{\bf 4. Resolution of the divergencies.} Because of the poles in
finite-temperature form factors of twist fields, the
representation (\ref{ftffexp}), as it stands, suffers from
divergencies at colliding rapidities, when they are associated to
opposite values of $\ep$. Using the link, provided by
(\ref{relftcyl}), between this representation and the form factor
expansion on the circle, we will describe the correct prescription
that makes the integrals in (\ref{ftffexp}) finite. This point was
left slightly unclear in the proposal \cite{LeclairLSS96}, were a
similar problem arose. Such divergencies at colliding rapidities,
however, were studied more carefully, in a related context, in the
recent paper \cite{AltshulerT05} using ideas similar to those
presented here.

Some of the ideas that we will present have an easy generalization
to interacting integrable models; this raises hopes that
finite-temperature form factors can also be evaluated in such
models. Work in this direction is in progress.

The paper is organized as follows. In Section 2, we recall the
quantization of the Majorana theory on the line and on the circle.
In Section 3, we introduce the space of ``finite-temperature
state'' and explain how finite-temperature form factors give
large-distance expansions of correlation functions. In Section 4,
we give an argument from basic principles justifying the relation
between finite-temperature form factors and form factors on the
circle. In Section 5, we study finite-temperature form factors of
non-interacting fields. In Section 6, we derive the
Riemann-Hilbert problem for finite-temperature form factors of
twist fields. In Section 7, we calculate these finite-temperature
form factors. Finally, we conclude with a series of open problems.

\sect{Quantization of the Majorana theory on the line and on the
circle}

In the free Majorana theory with mass $m$ quantized on the line,
fermion operators evolved in Euclidean time $\tau$ are given by:
\beqa
    \psi(x,\tau) &=& \frc12\sqrt{\frc{m}{\pi}}\,\int d\theta\,
    e^{\theta/2} \lt(
    a(\theta)\,e^{ip_\theta x - E_\theta\tau} +
    a^\dag(\theta)\,e^{-ip_\theta x + E_\theta\tau}
    \rt) \n
    \b\psi(x,\tau) &=& -\frc{i}2\,\sqrt{\frc{m}\pi}\,\int
    d\theta\,
    e^{-\theta/2} \lt(
    a(\theta)\,e^{ip_\theta x - E_\theta\tau} -
    a^\dag(\theta)\,e^{-ip_\theta x + E_\theta\tau}
    \rt)
\eeqa
where the mode operators $a(\theta)$ and their Hermitian conjugate
$a^\dag(\theta)$ satisfy the canonical anti-commutation relations
\beq\label{algtheta}
    \{a^\dag(\theta),a(\theta')\} = \delta(\theta-\theta')
\eeq
(other anti-commutators vanishing) and where
\beqa
    p_\theta &=& m\sinh\theta~, \n E_\theta &=& m\cosh\theta~.\no
\eeqa
The fermion operators satisfy the equations of motion
\beqa\label{edm}
    \b\p\psi(x,\tau) &\equiv& \frc12\lt(\p_x + i\,\p_\tau\rt) \psi =
    \frc{m}2 \b\psi \n
    \p\b\psi(x,\tau) &\equiv& \frc12\lt(\p_x - i\,\p_\tau\rt) \b\psi =
    \frc{m}2 \psi
\eeqa
and obey the equal-time anti-commutation relations
\beq\label{equaltime}
    \{\psi(x),\psi(x')\} = \delta(x-x')~,\quad
    \{\b\psi(x),\b\psi(x')\} = \delta(x-x')~,
\eeq
which is simple to derive from the representation
\beq
    \delta(x) = \frc1{2\pi} \int dp\,e^{ipx}
\eeq
of the delta-function. The Hilbert space $\cal H$ is simply the
Fock space over the algebra (\ref{algtheta}) with vacuum vector
$|\vac\ket$ defined by $a(\theta)|\vac\ket=0$. Vectors in $\cal H$
will be denoted by
\beq
    |\theta_1,\ldots,\theta_k\ket = a^\dag(\theta_1)\cdots
    a^\dag(\theta_k)|\vac\ket~.
\eeq
A basis is formed by taking, for instance,
$\theta_1>\cdots>\theta_k$. The Hamiltonian is given by
\beq
    H = \int_{-\infty}^{\infty}
    d\theta\,m\cosh\theta\,a^\dag(\theta) a(\theta)
\eeq
and has the property of being bounded from below on $\cal H$ and
of generating time translations:
\beq\label{Hpsi}
    [H,\psi(x,\tau)] = \frc{\p}{\p \tau} \psi(x,\tau)~,
    \quad [H,\bar\psi(x,\tau)] = \frc{\p}{\p \tau}
    \bar\psi(x,\tau)~.
\eeq
For future reference, note that the leading terms of the OPE's
$\psi(x,\tau)\psi(0,0)$ and $\bar\psi(x,\tau)\bar\psi(0,0)$ are
given by
\beq\label{OPEpsipsi}
    \psi(x,\tau)\psi(0,0) \sim \frc{i}{2\pi(x+i\tau)}~,
    \quad
    \b\psi(x,\tau)\b\psi(0,0) \sim -
    \frc{i}{2\pi(x-i\tau)}~.
\eeq
This normalization implies that the Majorana fermions are real,
and is different from the usual normalization used in conformal
field theory.

On the other hand, in the same theory quantized on the circle of
circumference $L$, with anti-periodic conditions on the fermion
fields, the fermion operators evolved in Euclidean time $\tau$
are:
\beqa
    \psi_L(x,\tau) &=& \frc1{\sqrt{2L}}
    \,\sum_{n\in\Z+\frc12}\frc{e^{\alpha_n/2}}{\sqrt{\cosh\alpha_n}}\,
    \lt(
    a_n\,e^{ip_n x - E_n\tau} +
    a^\dag_n\,e^{-ip_n x + E_n\tau}
    \rt) \n
    \b\psi_L(x,\tau) &=& -\frc{i}{\sqrt{2L}}
    \,\sum_{n\in\Z+\frc12}\frc{e^{-\alpha_n/2}}{\sqrt{\cosh\alpha_n}}\,
    \lt(
    a_n\,e^{ip_n x - E_n\tau} -
    a^\dag_n\,e^{-ip_n x + E_n\tau}
    \rt)
\eeqa
where the discrete mode operators $a_n$ and their Hermitian
conjugate $a^\dag_n$ satisfy the canonical anti-commutation
relations
\beq\label{algn}
    \{a^\dag_n,a_{n'}\} = \delta_{n,n'}
\eeq
(other anti-commutators vanishing) and where
\beqa
    p_n &=& m\sinh\alpha_n = \frc{2\pi n}L \quad (n\in\Z+\frc12) ~, \\  E_n &=&
    m\cosh\alpha_n~.\no
\eeqa
The fermion operators satisfy the equations of motion (\ref{edm})
as well as the equal-time anti-commutation relations
(\ref{equaltime}) (with the replacement $\psi\mapsto\psi_L$ and
$\b\psi\mapsto\b\psi_L$); the latter is simple to derive from the
representation
\beq
    \delta(x) = \frc1{L} \sum_{n\in\Z+\frc12}e^{ip_n x}
\eeq
of the delta-function, valid on the space of antiperiodic
functions on an interval of length $L$. The Hilbert space ${\cal
H}_L$ is simply the Fock space over the algebra (\ref{algn}) with
vacuum vector $|\vac\ket_L$ defined by $a_n|\vac\ket_L=0$. Vectors
in ${\cal H}_L$ will be denoted by
\beq
    |n_1,\ldots,n_k\ket_L = a^\dag_{n_1}\cdots
    a^\dag_{n_k}|\vac\ket_L~.
\eeq
A basis is formed by taking, for instance, $n_1>\cdots>n_k$. The
Hamiltonian is given by
\beq
    H_L = \sum_{n\in\Z+\frc12} m\cosh\alpha_n\,a^\dag_n a_n
\eeq
and has the property of being bounded from below on ${\cal H}_L$
and of generating time translations:
\beq
    [H_L,\psi_L(x,\tau)] = \frc{\p}{\p \tau} \psi_L(x,\tau)~,
    \quad [H_L,\bar\psi_L(x,\tau)] = \frc{\p}{\p \tau}
    \bar\psi_L(x,\tau)~.
\eeq
The leading terms of the OPE $\psi_L(x,\tau)\psi_L(0,0)$ and
$\bar\psi_L(x,\tau)\bar\psi_L(0,0)$ are the same as in
(\ref{OPEpsipsi}).

\sect{The space of ``finite-temperature states'' and
large-distance expansion of correlation functions}

\label{sectdefftff}

We are interested in calculating correlation functions in the
theory with Hamiltonian $H$ at finite temperature $1/L$. They are
given by ratios of traces on $\cal H$ as in (\ref{ftcorr}). In
particular, the relation (\ref{cylcorr}) holds, where for
non-interacting fields (normal-ordered products of fermion fields
and of their space derivatives), the operator $\Or_L$ can be
defined by first writing the regularization of $\Or$ in terms of a
product of fermion operators $\psi,\,\b\psi$ at different points
(without normal ordering), then by replacing all fermion operators
by their corresponding forms on the circle:
$\psi\mapsto\psi_L,\;\b\psi\mapsto\b\psi_L$.

The space of ``finite-temperature states'' that we will use is in
fact well-known in the literature: it is the Liouville space
\cite{vanHove86}, or the space of operators of the theory,
$\End({\cal H})$. As recalled in the introduction, the study of
finite-temperature quantum field theory using the Liouville space
goes under the name of thermo-field dynamics. It is not our
intention to go into any detail of this field of study; we will
recall only basic principles, in a slightly different, but
equivalent, formulation, then develop certain parts for our
purposes.

We will denote by $\ft$ the (completion of the)
infinite-dimensional subspace of $\End({\cal H})$ spanned by
products of any finite number of mode operators
$a(\theta),\;a^\dag(\theta)$ at different rapidities. This
subspace contains local non-interacting fields of the Majorana
theory. The vacuum in $\ft$ is defined by the identity on $\cal
H$:
\beq
    |\vac\ket^\ft \; \equiv \; {\bf 1}_{\cal H}~,
\eeq
and a complete basis of states is given by
\beq\label{basis}
    |\theta_1,\ldots,\theta_k\ket_{\ep_1,\ldots,\ep_k}^{\ft}
    \equiv \prod_{j=1}^k \lt(1+e^{-\ep_jLE_{\theta_j}}\rt) a^{\ep_1}(\theta_1)\cdots
    a^{\ep_k}(\theta_k)~,\qquad
    \theta_1>\theta_2>\cdots>\theta_k
\eeq
where $\ep_j$ are signs $(\pm)$, which will be called ``charge''
of the ``particle'' of rapidity $\theta_j$. In fact, the sign
$\ep_j$ is associated to the creation from the thermal bath, if it
is positive, or to the absorption by the thermal bath, if it is
negative, of a particle of rapidity $\theta_j$. Note that in
(\ref{basis}), we introduced the factor $\prod_{j=1}^k
\lt(1+e^{-\ep_jLE_{\theta_j}}\rt)$; this just specifies the
normalization of the states, but will play an important role in
the following. The inner product on $\ft$ is defined by
\beq\label{inner}
    \bra u|v\ket \equiv \braL U^\dag
    V\ketL ~,\quad \mbox{for } |u\ket \equiv
    U,\;|v\ket \equiv V~.
\eeq
In fact, as usual, it will be convenient to define states with
other orderings of the rapidities by
\beq\label{allorderings}
    |\theta_1,\ldots,\theta_k\ket_{\ep_1,\ldots,\ep_k}^\ft =
    \lt\{\ba{ll} \displaystyle
        (-1)^P
        |\theta_{P(1)},\ldots,\theta_{P(k)}\ket_{\ep_{P(1)},\ldots,\ep_{P(k)}}^\ft
        & (\theta_i\neq\theta_j \forall i\neq j)\\
        0 & (\exists i\neq j \;|\; \theta_i=\theta_j)~.
        \ea \rt.
\eeq
Here $P$ is a permutation of the $k$ first integers such that
$\theta_{P(1)} > \cdots > \theta_{P(k)}$, and $(-1)^P$ is the sign
of the permutation. Operators on $\cal H$ can also be mapped to
operators on $\ft$ by implementing their action on $\ft$ through
their left-action or their right-action on endomorphisms of $\cal
H$. It will be convenient for our purposes to concentrate solely
on their left-action:
\beq\label{leftaction}
    \Or \in \End({\cal H}) \mapsto \Or^\ft \in \End(\ft) \equiv
    \mbox{ left-action of } \Or \mbox{ on } \End({\cal H})~.
\eeq

Using the cyclic property of traces, it is easy to calculate the
following trace:
\beqa
    \braL a(\theta)a^\dag(\theta') \ketL &=&
    e^{LE_\theta}
    \braL a^\dag(\theta')a(\theta)\ketL
    \n
    &=&
    e^{LE_\theta} \lt(
    -\braL a(\theta)a^\dag(\theta')\ketL
    + \delta(\theta-\theta')\rt) \n
    &=& \frc{\delta(\theta-\theta')}{1+e^{-LE_\theta}}~.
\label{traadag}
\eeqa
Similarly,
\beq\label{tradaga}
    \braL a^\dag(\theta)a(\theta')\ketL
    = \frc{\delta(\theta-\theta')}{1+e^{LE_\theta}}
\eeq
and we also have
\beq\label{traaa}
    \braL a(\theta)a(\theta')\ketL
    =
    \braL a^\dag(\theta)a^\dag(\theta')\ketL
    =0~.
\eeq
From these results and from using Wick's theorem (which applies
for traces of products of free modes), it is easily seen that
\beq
    {\ }_{\ep_1,\ldots,\ep_k}\hspace{-4mm}{\
    }^\ft\bra\theta_1,\ldots,\theta_k
    |\theta_1',\ldots,\theta_k'\ket_{\ep_1',\ldots,\ep_k'}^\ft
    = \delta(\theta_1-\theta_1')\cdots\delta(\theta_k-\theta_k')~
    \delta_{\ep_1,\ep_1'}\cdots\delta_{\ep_k,\ep_k'}
    \prod_{j=1}^k\lt(1+e^{-\ep_jLE_{\theta_j}}\rt)
\eeq
if $\theta_1>\cdots>\theta_k$ and $\theta_1'>\cdots>\theta_k'$,
and that other inner products with such an ordering of rapidities
are zero.

From these definitions, we see that finite-temperature expectation
values of operators on $\cal H$ are vacuum expectation values on
$\ft$:
\beq
    \braL \Or(x,\tau) \cdots \ketL = {\ }^\ft\bra\vac| \Or^\ft(x,\tau) \cdots |\vac\ket^\ft~.
\eeq
Hence, using the resolution of the identity on $\ft$:
\beq\label{resid}
    {\bf 1}_{\ft} = \sum_{k=0}^\infty \sum_{\ep_1,\ldots,\ep_k}
    \int \frc{d\theta_1\cdots d\theta_k}{k!\prod_{j=1}^k
    \lt(1+e^{-\ep_j
    LE_{\theta_j}}\rt)}\;
    |\theta_1,\ldots,\theta_k\ket_{\ep_1,\ldots,\ep_k}^\ft ~
    {\ }_{\ep_1,\ldots,\ep_k}\hspace{-4mm}{\
    }^\ft\bra\theta_1,\ldots,\theta_k|~,
\eeq
two-point functions can be expanded as in (\ref{ftffexp}), where
we define {\em finite-temperature form factors} as
\beq\label{ftffdef}
    f^{\Or}_{\ep_1,\ldots,\ep_k}(\theta_1,\ldots,\theta_k;L) =
    {\ }^\ft\bra\vac|\Or^\ft(0,0)|\theta_1,\ldots,\theta_k\ket_{\ep_1,\ldots,\ep_k}^\ft~.
\eeq
Of course, multi-point correlation functions can be expanded in
similar ways. Note that in obtaining (\ref{ftffexp}), we used the
symmetries under translations (along the cylinder and around the
cylinder) in order to bring out the $x$ and $\tau$ dependence of
$\Or(x,\tau)$ in exponential factors.

It is easy to check, using the cyclic property of traces, that the
finite-temperature form factors can be written in the form
(\ref{ftffcac}). This will be useful in understanding why these
matrix elements are related to form factors on the circle.

It is a simple matter to observe that in the zero-temperature
limit $L\to\infty$, the finite-temperature form factors reproduce
the zero-temperature form factors:
\beq\ba{r}
    \lim_{L\to\infty}
    f_{+,\ldots,+,-,\ldots,-}^{\Or}(\theta_1,\ldots,\theta_{k_+},\theta_{k_++1},\ldots,\theta_k;L)
    =
    \bra\theta_{k},\ldots,\theta_{k_++1}|\Or(0,0)|\theta_1,\ldots,\theta_{k_+}\ket
    \z
    (\theta_i\neq\theta_j\;\forall\;i\in\{1,\ldots,k_+\},\,j\in\{k_++1,\ldots,N\})
\ea
\eeq
where there are $k_+$ positive charges $(+)$'s in the indices of
$f^{\Or}$. For other orderings of the charges (and of the
associated rapidities), we get extra minus signs. The
finite-temperature form factor expansion (\ref{ftffexp}), in the
limit $L\to\infty$, then reproduces the usual form factor
expansion, since only the form factors with $\ep_j=+$ for all
$j$'s remain.

It is also a simple matter to observe that more general matrix
elements in the space $\ft$ are simply related to the
finite-temperature form factors (\ref{ftffdef}):
\beq\label{invcharge}\ba{r}
    f_{\ep_1,\ldots,\ep_k}^{\Or}(\theta_1,\ldots,\theta_{k_+},\theta_{k_++1},\ldots,\theta_k;L)
    =
    {\ }_{-\ep_k,\ldots,-\ep_{k_++1}}{\ }^{\hspace{-4mm} \ft}
    \bra\theta_{k},\ldots,\theta_{k_++1}|\Or^\ft(0,0)
        |\theta_1,\ldots,\theta_{k_+}\ket_{\ep_1,\ldots,\ep_{k_+}}^\ft
    \z
    (\theta_i\neq\theta_j\;\forall\;i\in\{1,\ldots,k_+\},\,j\in\{k_++1,\ldots,k\})~.
\ea
\eeq

Let us make two comments. First, note that expression
(\ref{ftffexp}), as it stands, is not a practical large-distance
expansion. It is more convenient, as in the zero-temperature case,
to analytically continue to complex values of $\theta_j$'s, in
such a way that the argument of the exponential,
$i\sum_{j=1}^k\ep_jp_{\theta_j}x$, acquires a real part that
becomes large and negative at large $|\theta_j|$ for any $j$. This
is optimal, if $x>0$, at $\theta_j\mapsto \theta_j + i\ep_j\pi/2$.
However, in contrast to the zero-temperature case, the theory is
not invariant under rotation, hence this shift of rapidities does
not keep the form factors invariant. We will see how this shift is
related to a form factor expansion on the circle.

Second, the integrals in the expansion (\ref{ftffexp}) may suffer
from singularities at colliding rapidities in their integrands. As
we will see, such singularities arise in finite-temperature form
factors of interacting fields when rapidity variables are
associated to opposite charges. We need a prescription to define
properly the finite-temperature form factor expansion in such a
case. The correct prescription is the following: Consider the
difference $x$ between the positions of the first and of the
second operator involved to be {\em positive}; then, we simply
shift slightly the integration lines towards the positive or
negative imaginary direction in rapidity planes, in such a way
that the integration stays convergent at $\tau=0$ and that the
poles are avoided. More precisely, the integrals on
$\theta_1,\ldots,\theta_k$ in (\ref{ftffexp}) should be on
contours parallel to the real line with ${\rm Im}(\theta_j) =
\ep_j 0^+$. We verify that this prescription is correct, by
showing independently, in Appendices \ref{appProofftcyl} and
\ref{appProofexp}, the formula (\ref{relftcyl}), and the fact that
our prescription makes the finite-temperature form factor
expansion (\ref{ftffexp}) and the form factor expansion on the
circle (\ref{cylffexp}) equivalent.

{\bf A note on Ramond and Neveu-Schwartz sectors.}

Because of the relation (\ref{KMS}), fermion operators satisfy
anti-periodic conditions in Euclidean time under the trace. The
trace (\ref{ftcorr}) with insertions of fermion fields then
corresponds to correlation functions on the cylinder with
anti-periodic conditions for the fermion fields around the
cylinder, that is, in the NS sector in the language of the
quantization on the circle.

When twist fields are inserted inside the trace, the situation is
slightly different. Twist fields are end-points of branch cuts (or
line defects), and on the cylinder, a semi-infinite branch cut can
be in two directions. The definition of operators corresponding to
twist fields is recalled in Appendix \ref{appTwist}: we denote two
types of operators corresponding to any twist field $\Or$ by
$\Or_+$ and $\Or_-$, with branch cut towards the positive and
negative $x$ direction along the cylinder, respectively. We call
these operators ``right-twist'' and ``left-twist'' operators,
respectively.

Finite-temperature correlation functions (\ref{ftcorr}) with
insertion of only one twist fields (and any number of fermion
fields) correspond then to correlation functions on the cylinder
in a mixed sector, the precise mixed sector depending on which
operator (right-twist or left-twist) is taken. The fermion fields
have periodic conditions around the cylinder (R sector) where the
branch cut stands, and anti-periodic conditions (NS sector)
elsewhere. This property is essential in deriving the
Riemann-Hilbert problem for finite-temperature form factors in
Section \ref{sectRiemann}. Correspondingly, our formula
(\ref{ftffexp}) can be used with any one of $\Or_1$ and $\Or_2$ a twist field,
and in both cases, the branch cut can be in either direction.

Finite-temperature correlation functions with insertion of two
twist fields can correspond to correlation functions on the
cylinder in a pure NS sector or in a pure R sector, again
depending on which operators are taken under the trace. Our
formula (\ref{ftffexp}), however, can only be used, in this case,
to give an expansion of correlation functions in the R sector.
This can be interpreted by the fact that in each
finite-temperature matrix element of a twist field, there is no
other twist field to provide another end-point of the branch cut,
so that it must go to infinity along the cylinder, hence it
changes the natural NS sector of the trace to an R sector at
infinity. This can be seen technically by analyzing our
Riemann-Hilbert problem given in Section \ref{sectRiemann}, and by
doing a contour-deformation analysis of the expansion
(\ref{ftffexp}), similarly to what is done in Appendix
\ref{appProofexp}. Correspondingly, then, the expansion
(\ref{ftffexp}) is valid, if both $\Or_1$ and $\Or_2$ are twist
fields, only when they have branch cuts in opposite directions.

In the same spirit, note that formula (\ref{relftcyl}) for
calculating form factors on the circle from finite-temperature
form factors, explained in the next section, can only be used with
the vacuum (in the quantization on the circle) in the R sector,
and excited states in the NS sector.

However, it can be useful to note that one could ``twist'' the
construction of this section by defining the inner product as
traces with {\em periodic conditions} on the fermion fields:
$$\bra u|v\ket \equiv \braL \sigma_+(-\infty,0) U^\dag V
\ketL/\braL\sigma_+\ketL~,$$ where the operator $\sigma_+$ is the
operator associated to a primary twist field of non-zero
finite-temperature average, with a branch cut on the right. The
resulting finite-temperature correlation functions of fermion
fields will be in the R sector. With such a definition, the
formula (\ref{ftffcac}) is still valid, but the normalization
factors $\prod_{j=1}^k \lt(1+e^{-\ep_jLE_{\theta_j}}\rt)$
introduced above should be replaced by $\prod_{j=1}^k
\lt(1-e^{-\ep_jLE_{\theta_j}}\rt)$, and a similar replacement
should be made in the finite-temperature form factor expansion
(\ref{ftffexp}). Following a similar reasoning as above, the
resulting expansion for the two-point function of twist fields
will be valid, in this construction, only in the NS sector. The
finite-temperature form factors obtained in this construction can
then be used to calculate form factors on the circle with excited
states in the R sector and vacuum in the NS sector, with a formula
similar to (\ref{relftcyl}). Note that starting from the form
factor expansion on the circle, a representation similar to
(\ref{ftffexp}) (more precisely, after some approximations, to
(\ref{ftffexpLLSS})) was obtained in \cite{AltshulerT05} for
two-point functions of order fields, with the factors
$\prod_{j=1}^k\lt(1+e^{-\ep_j LE_{\theta_j}}\rt)$ replaced by
$\prod_{j=1}^k\lt(1-e^{-\ep_j mE_{\theta_j}}\rt)$. The
construction just outlined in this paragraph explains this
phenomenon. In the rest of the paper, though, we will not consider
this construction.

\sect{From finite-temperature form factors to form factors on the
circle}

\ssect{General idea}

Observe that the two equivalent expressions (\ref{ftcorr}) and
(\ref{cylcorr}) for correlation functions at finite temperature,
or on a cylinder, result from two different quantization schemes
for the same theory: in the first, the equal-time slices are lines
along the cylinder, whereas in the second, they are circles around
it.

Observe also that the finite-temperature form factor expansion of
two-point functions (\ref{ftffexp}) could be made into a form
factor expansion on the circle (\ref{cylffexp}), if
finite-temperature form factors have appropriate properties, by
deforming the contours in rapidity space and taking the residues
of the poles of the measure.

There are more precise observations that allow us to infer that
form factors on the circle can be obtained from finite-temperature
form factors. Let us develop two of them below.

First, consider a massive free theory on the Poincar\'e disk (the
Dirac theory on the Poincar\'e disk was studied in \cite{Doyon03},
and we present here ideas exposed there), and consider a
quantization scheme where equal-time slices are the orbits of one
of its non-compact isometry (that is, the momentum operator
generates isometry transformations $\cal I$ and its spectrum is
continuous). In this quantization scheme, consider the set of
states $|P\ket$ that diagonalize the momentum operator, with real
continuous eigenvalues $P$. In fact, the states can be described
by a set of momenta $\{p\}$ and a set of discrete variables
(particle momenta and particle types).

Now consider matrix elements of local spinless operators in this
set of states, of the form $\bra\vac|\Or|\{p\}\ket$, and see them
as analytically continued functions of momenta $\{p\}$. Choose a normalization
of the states in such a way that these matrix elements are entire
functions of the momenta $\{p\}$'s. With this choice, there is a
measure involved in the resolution of the identity: ${\bf 1} =
\sum_k \int [dp]\,\rho(\{p\}) |\{p\}\ket\bra \{p\}|$ where $k$ is
the number of particles. The singularity structure of
$\rho(\{p\})$ then describes the energy spectrum of the theory in
the quantization scheme where the roles of the momentum and the
Hamiltonian are exchanged: the isometry $\cal I$ is generated by
the Hamiltonian. In particular, poles at purely imaginary values
$p=\pm iE$ correspond to single-particle discrete eigenvalues $E$
of the Hamiltonian in that scheme. Also, the matrix elements of
$\Or$ in the ``exchanged'' quantization scheme, of the form ${\
}' \bra\vac|\Or|\{E\}\ket'$ where $|\{E\}\ket'$ is an
energy eigenvalue, are proportional to
$\sqrt{\Res(\rho(\{p\}))}\,\bra\vac|\Or|\{p\}\ket$ at the values
$\{p=\pm iE\}$.

The second observation is a similar one, but this time for
interacting relativistic models on flat space. Consider for
simplicity such a model with only one particle type (one mass).
Consider a set of momentum eigenstates, parametrized by rapidity
variables $\{\theta\}$. Choose a normalization of the states such
that the form factors of any local field are entire functions of
the total rapidity when rapidity differences are fixed, and such
that the measure involved in the resolution of the identity ${\bf
1} = \sum_k \int [d\theta] \rho(\{\theta\})
|\{\theta\}\ket\bra\{\theta\}|$, where $k$ is the number of
particles, factorizes into a product of measures for the
individual rapidities of each particle. This is the usual choice:
the dependence of form factors of local fields on the total
rapidity is the entire function $e^{s\sum_j \theta_j}$, where $s$
is the spin of the field; and the measure is $\rho(\{\theta\}) =
1$. To clarify the idea, the resolution of the identity can be
written as integration over momentum variables with an appropriate
relativistically invariant measure: ${\bf 1} = \sum_k \int
\lt[\frc{dp}{\sqrt{p^2+m^2}}\rt] |\{\theta\}\ket\bra\{\theta\}|$.
The measure has branch cuts from $p=\pm im$ towards $\pm i\infty$:
these describe, again, the energy spectrum of the theory where the
roles of the momentum and of the Hamiltonian are exchanged. Of
course here, by relativistic invariance, this exchange gives the
same Hilbert space. It can be implemented by the shift
$\theta\mapsto\theta+ i\pi/2$ of rapidity variables, and the form
factors are invariant, up to a phase due to the spin, under such a
shift of rapidities.

Similarly, it is not hard to relate analytical continuation of
special traces in angular quantization to the spectrum and matrix
elements of local operators in radial quantization. Adapting these
observations to the case of the cylinder, the main idea for
obtaining form factors on the circle from finite-temperature form
factors in the free Majorana theory can be expressed in the
following (schematic) steps:
\begin{itemize}
\item Consider the Liouville space $\ft$, where the trace
\[
    \frc{\Tr\lt(e^{-LH}
    \Or(x,0)\Or(0,0)\rt)}{\Tr\lt(e^{-LH}\rt)}
\]
is a vacuum expectation value and where eigenvalues of the
momentum operator are described by continuous variables, the
rapidities $\theta_j$;
\item Find a measure $\rho(\{\theta\})$ in
\[
    {\bf 1}_\ft = \sum_{k=0}^\infty \sum_{\{\ep\}}\int \frc{\{d\theta\}}{k!} \rho(\{\theta\})
    |\{\theta\}\ket^\ft_{\{\ep\}} {\ }^\ft_{\{\ep\}}\bra\{\theta\}|
\]
such that matrix elements of local fields $\Psi$ which are also
local with respect to the fundamental field $\psi$, $ {\ }^\ft
\bra\vac|\Psi(0,0)|\{\theta\}\ket^\ft_{\{\ep\}}$, are entire
functions of the rapidities $\{\theta\}$;
\item Calculate form factors on the circle by analytical
continuation in the rapidity variables to the positions $\alpha_n+
i\pi/2$ of the poles of the measure $\rho$:
\[
    {\ }_L\bra\vac|\Psi_L(0,0)|\{n\}\ket_L \propto
    \sqrt{{\rm Res}\,\rho}\,{\
    }^\ft\bra\vac|\Psi(0,0)|\{\alpha_n\pm
    i\pi/2\}\ket^\ft_{\{\pm\}}~.
\]
\end{itemize}
In the last step, $\alpha_n$ are defined in (\ref{defalphan}). In
the second step, the requirement that the fields $\Psi$ be local
with respect to the fundamental field $\psi$ comes from the fact
that rotation is not a symmetry on the cylinder.
Finite-temperature form factors of twist fields (which are
essentially the end-point of linear defects) can be expected to
have a more complicated pole structure.

More than relating finite-temperature form factors to form factors
on the circle, the steps above tell us that an appropriate choice
of the measure is related to a simple analytical structure of
finite-temperature form factors.

It is hopefully possible to generalize some of these requirements
to interacting integrable models, but this will not be pursued
further here.

\ssect{Implementation of the general idea}

For the massive Majorana theory, the first step was described in
the previous section.

We now show that the second step is realized by the definition
(\ref{ftffdef}), due to the representation (\ref{ftffcac}).
Consider a complete basis $\{\Psi_j\}$ of local operators that are
also local with respect to the fundamental field $\psi$; these are
non-interacting operators, for instance normal-ordered products of
fermion operators and of their space derivatives, with
coefficients that may contain integer powers of the mass $m$ (the
mass $m$ has scaling dimension 1 and spin 0). We consider the
basis $\{\Psi_j\}$ to be for the linear space of operators on the
field of complex numbers without dependence on $m$. We can take
the operators $\Psi_j$ to have well-defined spins $s_j$ and
scaling dimensions $d_j$. With the fermion operators $\psi$ and
$\bar\psi$, they satisfy simple equal-time commutation (or
anti-commutation) relations:
\beqa
    [\psi(x),\Psi_i(x')] &=& \sum_{j;\;d_j=0,\ldots,d_i-\frc12}
    c_{i}^{j}
    \;\Psi_j(x')\;
    \delta^{(d_i-d_j-\frc12)}(x-x') \n \label{localcomm}
{}    [\b\psi(x),\Psi_i(x')] &=&
    \sum_{j;\;d_j=0,\ldots,d_i-\frc12}
    \b{c}_{i}^{j} \;\Psi_j(x')\;
    \delta^{(d_i-d_j-\frc12)}(x-x')~.
\eeqa
Here the symbol $[\cdot,\cdot]$ is the commutator if $s_j$ is an
integer or the anti-commutator if $s_j$ is a half-integer (the
spin $s_j$ determines the statistics of the operator
$\Psi_{j}(x')$). The sum is of course finite, because there are no
fields of negative dimension. The coefficients
$c_{i}^{j},\;\b{c}_{i}^{j}$ do not depend on $m$. The fact that
$\Psi_j$ have well-defined spin and dimension imposes constraints
on the coefficients $c_{i}^{j},\;\b{c}_{i}^{j}$, but we will not
use these constraints here.

The operators $a(\theta)$ and $a^\dag(\theta)$ can be decomposed
in terms of the local operators $\psi(x)$ and $\b\psi(x)$:
\beq
    a^\ep(\theta) = \frc12\sqrt{\frc{m}\pi}\,\int_{-\infty}^{\infty} dx\, e^{\ep ip_\theta x}\,
        (e^{\theta/2}\psi(x)-\ep ie^{-\theta/2}\b\psi(x))~.
    \label{modespsi}
\eeq
The one-particle finite-temperature form factor of a field
$\Or_i(x)$ can then be written, from (\ref{ftffcac}),
\beqa
    f^{\Psi_i}_\ep(\theta;L) &=&
    \frc12\sqrt{\frc{m}\pi}\,\int_{-\infty}^{\infty} dx\, e^{\ep\,ip_\theta x}\,
        \lt(e^{\theta/2}\braL[\psi(x),\Psi_i(0)]\ketL-
        \ep\,ie^{-\theta/2}\braL[\b\psi(x),\Psi_i(0)]\ketL \rt) \n
    &=&
    \frc12\sqrt{\frc{m}\pi}\,\sum_{j;\;d_j=0}^{d_i-\frc12}
        \int_{-\infty}^{\infty} dx\, e^{\ep\,ip_\theta x}\,
        \lt(e^{\theta/2}c_i^j-
        \ep\,ie^{-\theta/2}\b{c}_i^j\rt)\braL \Psi_j(0)\ketL
        \delta^{(d_i-d_j-\frc12)}(x)\n
    &=&\label{ftff1plocal}
    \frc12\sqrt{\frc{m}\pi}\,\sum_{j;\;d_j=0}^{d_i-\frc12}
        (-\ep\,ip_\theta)^{d_i-d_j-\frc12}\,
        \lt(e^{\theta/2}c_i^j-
        \ep\,ie^{-\theta/2}\b{c}_i^j\rt) \braL \Psi_j(0)\ketL
\eeqa
Note that these finite-temperature form factors are entire
functions of the variable $\theta$.

Similarly, the multi-particle form-factors are given by
\beqa \label{ftffunint}
&&
    f^{\Psi_{j_{k+1}}}_{\ep_1,\ldots,\ep_k}(\theta_1,\ldots,\theta_k;L)
    = \\ && \qquad
    \lt(\frc12\sqrt{\frc{m}\pi}\rt)^k \sum_{j_1,\ldots,j_k}
    \prod_{l=1}^{k} \lt[
    \lt(-\ep_l\,ip_{\theta_l}\rt)^{d_{j_{l+1}}-d_{j_{l}}-\frc12}
    \lt(e^{\theta_l/2}c_{j_{l+1}}^{j_l} -
    \ep_{l}\,ie^{-\theta_l/2} \b{c}_{j_{l+1}}^{j_l} \rt) \rt]\;
    \braL \Psi_{j_1}(0)\ketL ~. \no
\eeqa
Again, they are entire functions of the variables $\theta_l$.
Hence, the measure appearing in the resolution of the identity
(\ref{resid}) fulfills the condition of the second step.

The third step then tells us how to evaluate form factors on the
circle. The precise relation can be obtained, for instance, by
calculating explicit examples for simple non-interacting fields.
It is given by (\ref{relftcyl}). A more direct derivation of this
relation is presented in Appendix \ref{appProofftcyl}. A proof
that the finite-temperature form factor expansion of two-point
functions (\ref{ftffexp}) is equivalent to the expansion in form
factors on the circle (\ref{cylffexp}) is presented in Appendix
\ref{appProofexp}, using properties of finite-temperature form
factors that are derived in the rest of this paper.

\sect{Finite-temperature form factors of non-interacting fields}

\label{sectftffunint}

\ssect{General properties and particular cases}

The expression (\ref{ftffunint}) gives a general formula for
finite-temperature form factors of local non-interacting
operators. In particular, it is an easy matter to verify the
property
\beq\label{crossingunint}
    f_{\ep_1,\ldots,\ep_j,\ldots,\ep_k}^{\Psi_i}(\theta_1,\ldots,\theta_j+i\pi,\ldots,\theta_k;L)
     = i
    f_{\ep_1,\ldots,-\ep_j,\ldots,\ep_k}^{\Psi_i}(\theta_1,\ldots,\theta_j,\ldots,\theta_k;L)~.
\eeq
This leads to the quasi-periodicity property
\beq\label{qpunint}
    f_{\ep_1,\ldots,\ep_j,\ldots,\ep_k}^{\Psi_i}(\theta_1,\ldots,\theta_j+2i\pi,\ldots,\theta_k;L)
     = -
    f_{\ep_1,\ldots,\ep_j,\ldots,\ep_k}^{\Psi_i}(\theta_1,\ldots,\theta_j,\ldots,\theta_k;L)~.
\eeq

Note that the same formula (\ref{ftffunint}) could be used, with
the replacement $\braL\Psi_j\ketL\mapsto \bra\Psi_j\ket$, for
calculating form factors at zero temperature; indeed, the
parameter $L$ is only involved in the expectation values
$\braL\Psi_j\ketL$. This suggests that finite-temperature form
factors of non-interacting operators are ``not so far'' from their
zero-temperature limit. For instance, the finite-temperature form
factors of the fundamental fermion operators are
\beq\label{ftfffermions}
    f^\psi_\pm(\theta;L) =
    \frc12\sqrt{\frc{m}\pi} e^{\theta/2}~,\qquad
    f^{\b\psi}_\pm(\theta;L) = \mp i
    \frc12\sqrt{\frc{m}\pi} e^{-\theta/2}~,
\eeq
which are temperature independent. On the other hand, for the
energy field (from the viewpoint of the Ising field theory)
$\varepsilon =i :\b\psi\psi:$, the two-particle finite-temperature
form factors are
\beq
    f^\varepsilon_{+,+}(\theta_1,\theta_2;L) = -f^\varepsilon_{-,-}(\theta_1,\theta_2;L)
    = \frc{m}{2\pi}
    \sinh\lt(\frc{\theta_1-\theta_2}2\rt)
\eeq
and
\beq
    f^\varepsilon_{+,-}(\theta_1,\theta_2;L) = -f^\varepsilon_{-,+}(\theta_1,\theta_2;L)
    = -\frc{m}{2\pi}
    \cosh\lt(\frc{\theta_1-\theta_2}2\rt)
\eeq
which agree with the zero-temperature form factors, but the
thermal expectation value is
\beq
    f^\varepsilon(-;L) = \braL \varepsilon\ketL = \frc{m}{\pi} \int_0^\infty
    \frc{d\theta}{1+e^{-mL\cosh\theta}}
\eeq
which is non-zero and $L$-dependent. Those in fact are the
zero-temperature form factors of the field $\varepsilon +
\braL\varepsilon\ketL {\bf 1}$. One could re-define the field
$\varepsilon$ by subtracting this thermal expectation value, but
for a more general non-interacting field, such a subtraction will
not bring all its finite-temperature form factors equal to its
zero-temperature form factors. This aspect was missing in
\cite{LeclairLSS96}. In general, fields that are normal-ordered
products of $k$ fermion fields have $k$-particle
finite-temperature form factors independent of temperature, but
generically have non-zero, temperature-dependent $j$-particle form
factors for $j<k$ ($j$ and $k$ being of the same parity). In order
to describe this, we introduce the concept of mixing, which is a
simple generalization of the description
$\varepsilon\mapsto\varepsilon + \braL\varepsilon\ketL {\bf 1}$ of
finite-temperature form factors of the energy field in terms of
its zero-temperature form factors.

\ssect{Mixing}

Consider a complete basis $\{\Psi_a\}$ of local non-interacting
operators on the field of polynomials in $m$ -- we will use
indices $a,b$ to label all elements of this set, instead of the
indices $i,j$ used in the previous section to label elements of a
different basis. The elements of this set of operators can be
taken to be in one to one correspondence with a basis of non-zero
operators in the free massless Majorana conformal field theory. A
convenient basis $\{\Psi_a\}$ is obtained by considering
normal-ordered products of fermion operators $\psi$ and $\b\psi$
and of their holomorphic or anti-holomorphic derivatives,
$\p^k\psi,\,\b\p^l\b\psi$, with coefficients that are independent
of the mass $m$. We will consider their holomorphic and
anti-holomorphic dimensions $\Delta_a,\,\b\Delta_a$, related as
usual to their scaling dimensions $d_a=\Delta_a+\b\Delta_a$ and
their spins $s_a=\Delta_a-\b\Delta_a$.

We will show that:
\beq\label{mixing}\ba{l}
    f_{+,\ldots,+,-,\ldots,-}^{\Psi_a}
        (\theta_1,\ldots,\theta_{k_+},\theta_{k_++1},\ldots,\theta_k;L)
    =
    \bra\theta_{k},\ldots,\theta_{k_++1}|\sum_b L^{d_b-d_a} M_a^b(m L) \Psi_b(0)
    |\theta_1,\ldots,\theta_{k_+}\ket
    \z
    \hfill (\theta_i\neq\theta_j\;\forall\;i\in\{1,\ldots,k_+\},\,j\in\{k_++1,\ldots,k\})
    \ea
\eeq
where there are $k_+$ positive charges $(+)$ in the indices of
$f^{\Psi_a}$. Here the mixing matrix $M_a^b(mL)$ mixes operators
$\Psi_b$ of lower dimension than that of $\Psi_a$, $d_b<d_a$, of
equal or lower associated conformal dimensions,
$\Delta_b\leq\Delta_a,\,\b\Delta_b\leq\b\Delta_a$, and of the same
statistics. The operator $\Psi_a$ is a descendant under the
fermion operator algebra of all operators $\Psi_b$; in other
words, the mixing occurs between $\Psi_a$ and some of its
ascendants of the same statistics. The sum over $b$ is finite.

One can see the space $\ft$ as a Fock space over the algebra of
modes $A_\ep(\theta)$ and of their Hermitian conjugate on $\ft$,
$A^\dag_\ep(\theta)$:
\beq
    \{A^\dag_\ep(\theta),A_{\ep'}(\theta')\} =
    (1+e^{-\ep LE_\theta})
    \,\delta(\theta-\theta')\,\delta_{\ep,\ep'}~,\quad
    \{A^\dag_\ep(\theta),A^\dag_{\ep'}(\theta')\} =
    \{A_\ep(\theta),A_{\ep'}(\theta')\} = 0~.
\eeq
The vacuum is defined by $A_\ep(\theta)|\vac\ket^\ft=0$, and
\beq
    |\theta_1,\ldots,\theta_N\ket_{\ep_1,\ldots,\ep_N}^\ft =
    A_{\ep_1}^\dag(\theta_1)\cdots
    A_{\ep_N}^\dag(\theta_N)|\vac\ket^\ft~.
\eeq
In particular, according to (\ref{leftaction}), the operators
$a(\theta)$ and $a^\dag(\theta)$ act on $\ft$ by
\beq
    [a^\ep(\theta)]^\ft = \frc{A_\ep^\dag(\theta)}{1+e^{-\ep LE_\theta}}
    + \frc{A_{-\ep}(\theta)}{1+e^{\ep LE_\theta}}~.
\eeq
One can immediately verify, for instance, that this representation
on $\ft$ gives $\{[a^\ep(\theta)]^\ft,[a^{\ep'}(\theta')]^\ft\} =
\delta(\theta-\theta')\,\delta_{\ep,\ep'}$, as it should.

Now consider an explicit set of local non-interacting operators
$\{\Psi_a\}$ given by the coefficients in the expansion around
$x_1=0,\ldots,x_{p}=0$ of the generating operator
\beq
    \Psi_p(x_1,\ldots,x_p) =\;
    :\psi_1(x_1,i\ep_1x_1)\cdots\psi_p(x_p,i\ep_px_p):
\eeq
where each element of the set of fields $\{\psi_j,\,1\le j\le p\}$
is one of $\psi$ or $\b\psi$, and where $\ep_j=-1$ if
$\psi_j=\psi$ and $\ep_j=1$ if $\psi_j=\b\psi$. According to
(\ref{leftaction}), the corresponding operator acting on $\ft$ by
left action is denoted by $\Psi_{p,q}^\ft(x_1,\ldots,x_{p+q})$.
The normal-ordering operation $:\cdot:$ on ${\cal H}$ brings the
operators $a(\theta)$ to the right of the operators
$a^\dag(\theta)$. This normal ordering operation can also be
defined on operators acting on $\ft$ by bringing operators
$A_+(\theta)$ and $A^\dag_-(\theta)$ to the right of
$A_+^\dag(\theta)$ and $A_-(\theta)$, so that we can write
\beq
    \Psi_p^\ft(x_1,\ldots,x_{p}) =\;
    :\psi_1^\ft(x_1,i\ep_1x_1)\cdots\psi_p^\ft(x_p,i\ep_px_p):~.
\eeq
We can also define a more natural normal-ordering operation on
$\ft$ which brings all operators $A_+(\theta)$ and $A_-(\theta)$
to the right of $A_+^\dag(\theta)$ and $A_-^\dag(\theta)$; let's
denote it by $:\cdot:_\ft$. We introduce the generating operator
\beq
    \tilde\Psi_p^\ft(x_1,\ldots,x_{p}) =\;
    :\psi_1(x_1,i\ep_1x_1)\cdots\psi_p(x_p,i\ep_px_p):_\ft \;=
    \; :\Psi_p^\ft(x_1,\ldots,x_p):_\ft~.
\eeq
By Wick's theorem, we have
\beqa
&&    \Psi_{p}^\ft(x_1,\ldots,x_{p}) + \sum_{m,n,\;m<n}
        (-1)^{m-n-1}C_{m,n}(x_m-x_n)
        \Psi_{p-2}^\ft(x_1,\ldots,\h{x}_m,\ldots,\h{x}_n,\ldots,x_{p})
        + \ldots \n
\label{Wick}        && \qquad = \\
&&    \tilde\Psi_{p}^\ft(x_1,\ldots,x_{p}) + \sum_{m,n,\;m<n}
        (-1)^{m-n-1}\tilde{C}_{m,n}(x_m-x_n)
        \tilde\Psi_{p-2}^\ft(x_1,\ldots,\h{x}_m,\ldots,\h{x}_n,\ldots,x_{p})
        + \ldots ~. \no
\eeqa
Here $C_{m,n}(x_m-x_n)$ is the vacuum expectation value on $\cal
H$ of the product of the fermion field at $x_m$ and the fermion
field at $x_n$; that is, $C_{m,n}(x_m-x_n) =
\bra\vac|\psi_m(x_m,i\ep_mx_m)\psi_n(x_n,i\ep_nx_n)|\vac\ket$.
Similarly, $\tilde{C}_{m,n}(x_m-x_n)$ is the vacuum expectation
value on $\ft$ of the product of the field at $x_m$ and the field
at $x_n$; that is, $\tilde{C}_{m,n}(x_m-x_n) = {\ }^\ft
\bra\vac|\psi_m^\ft(x_m,i\ep_mx_m)\psi_n^\ft(x_n,i\ep_nx_n)|\vac\ket^\ft
= \braL\psi_m(x_m,i\ep_mx_m)\psi_n(x_n,i\ep_nx_n)\ketL$. The dots
($\ldots$) indicate terms with more and more contractions, until
all fields are contracted, in the usual way.

By solving iteratively (\ref{Wick}), we can write
\beq\label{mixingbasis}
    \Psi_{p}^\ft(x_1,\ldots,x_{p}) =
    \tilde\Psi_{p}^\ft(x_1,\ldots,x_{p})+ \sum_{m,n,\,m<n}
    M_{m,n}(x_m-x_n)\tilde\Psi_{p-2}^\ft(x_1,\ldots,\h{x}_m,\ldots,\h{x}_n,\ldots,x_{p})
    + \ldots
\eeq
where $M_{m,n}(x_m-x_n) = (-1)^{m-n-1} (\t{C}_{m,n}(x_m-x_n) -
C_{m,n}(x_m-x_n))$. The dots ($\ldots$) represent terms containing
operators with decreasing total number of fields.

Consider the matrix element
\beq
    {\ }^\ft\bra\vac|\tilde\Psi_{p}^\ft(x_1,\ldots,x_{p})
        |\theta_1,\ldots,\theta_{p}\ket_{\ep_1,\ldots,\ep_{p}}^\ft~.
\eeq
Since there are as many rapidities in the state as there are
fermion operators in the operator of which we take the matrix
element, the normal ordering does not affect the result (there can
be no internal contractions). Hence we have
\beq
    {\ }^\ft\bra\vac|\tilde\Psi_{p}^\ft(x_1,\ldots,x_{p})
        |\theta_1,\ldots,\theta_{p}\ket_{\ep_1,\ldots,\ep_{p}}^\ft
    =
    {\ }^\ft\bra\vac|\Psi^\ft_{p}(x_1,\ldots,x_{p})
        |\theta_1,\ldots,\theta_{p}\ket_{\ep_1,\ldots,\ep_{p}}^\ft~.
\eeq
Moreover, by dimensional analysis, in the expression
(\ref{ftffunint}) for the right-hand side of the equation above,
only vacuum expectation values of operators of dimension 0 will
remain: the identity operator. Since the finite-temperature
expectation value of the identity operator is independent of the
temperature, and since the only $L$-dependence in
(\ref{ftffunint}) occurs in expectation values, we can specialize
the right-hand side of the previous equation to zero temperature
($L\to\infty$) without change in the equation. It is convenient at
this point to take, without loss of generality, the $p_+$ first
charges to be positive, and the rest to be negative, and to assume
$\theta_i\neq\theta_j\;\forall\;i\in\{1,\ldots,p_+\},\,j\in\{p_++1,\ldots,p\}$.
Then we find:
\beq\ba{l}
    {\ }^\ft\bra\vac|\tilde\Psi_{p}^\ft(x_1,\ldots,x_{p})
        |\theta_1,\ldots,\theta_{p_+},\theta_{p_++1},\ldots\theta_{p}
            \ket_{+,\ldots,+,-,\ldots,-}^\ft
    =\z\qquad\qquad\qquad\qquad\qquad\qquad\qquad
    \bra\theta_p,\ldots,\theta_{p_++1}|\Psi_{p}(x_1,\ldots,x_{p})
        |\theta_1,\ldots,\theta_{p_+}\ket~.
    \ea
\eeq
Now, all matrix elements of $\tilde\Psi_{p}^\ft(x_1,\ldots,x_{p})$
between the vacuum and states containing more or less than $p$
rapidities are zero because of the finite-temperature normal
ordering. Similarly, all zero-temperature matrix elements of
$\Psi_{p}(x_1,\ldots,x_{p})$ between states whose total number of
particle is more or less than $p$ are zero. Hence we can write in
general:
\beq\ba{l}
    {\ }^\ft\bra\vac|\tilde\Psi_{p}^\ft(x_1,\ldots,x_{p})
        |\theta_1,\ldots,\theta_{k_+},\theta_{k_++1},\ldots\theta_{k}
            \ket_{+,\ldots,+,-,\ldots,-}^\ft
    =\z\qquad\qquad\qquad\qquad\qquad\qquad\qquad
    \bra\theta_k,\ldots,\theta_{k_++1}|\Psi_{p}(x_1,\ldots,x_{p})
        |\theta_1,\ldots,\theta_{k_+}\ket
    \ea
\eeq
for any $k$ and any $k_+$. Using these equalities and the relation
(\ref{mixingbasis}), we have
\beqa
&&  f^{\Psi_{p}(x_1,\ldots,x_{p})}_{+,\ldots,+,-,\ldots,-}
        (\theta_1,\ldots,\theta_{k_+},\theta_{k_++1},\ldots\theta_{k};L)
         =
        \n
&&
    \bra\theta_k,\ldots,\theta_{k_++1}|\Psi_{p}(x_1,\ldots,x_{p})
        |\theta_1,\ldots,\theta_{k_+}\ket
        + \n
&&  \sum_{m<n}M_{m,n}(x_m-x_n)
    \bra\theta_k,\ldots,\theta_{k_++1}|\Psi_{p-2}(x_1,\ldots,\h{x}_m,\ldots,\h{x}_n,\ldots,x_{p})
        |\theta_1,\ldots,\theta_{k_+}\ket
        + \n
&&    \ldots~.
\eeqa
This is exactly of the form (\ref{mixing}), once expanded in
powers of $x_m$'s. From this calculation, we see that the elements
of the mixing matrix can be seen as coming from the difference
between the normal-ordering $:\cdot:_\ft$, which for local fields
on ${\cal H}$ is essentially a normal-ordering on the circle, and
the usual normal-ordering on the line $:\cdot:$.

It is possible to describe the mixing matrix in a more precise
fashion. Recall that the space $\ft$ contains local
non-interacting fields. The mixing matrix can be seen as an
operator acting on $\ft$, such that a state in $\ft$ corresponding
to the field $\Psi_a(0)$ is mapped to the state corresponding to
$\sum_{b} L^{d_b-d_a} M_a^b(mL) \Psi_b(0)$.

Consider the map $\Psi(0)\mapsto\tilde\Psi(0)$ that maps any
normal-ordered non-interacting operator $\Psi(0)\in\End({\cal H})$
to the operator $\tilde\Psi(0)\in\End({\cal H})$ such that
$\tilde\Psi^\ft(0) =\; :\Psi^\ft(0):_\ft$ (the finite-temperature
form factors of $\tilde\Psi(0)$ are equal to the zero-temperature
form factors of $\Psi(0)$, as described above). We saw above how
to construct $\tilde\Psi(0)$: it indeed exists, and it is unique.
Now consider the state $|\Psi\ket^\ft = \Psi^\ft(0)|\vac\ket^\ft$
as well as the state $|\tilde\Psi\ket^\ft =
\tilde\Psi^\ft(0)|\vac\ket^\ft$ in $\ft$. The mixing matrix
$M_a^b(mL)$ is defined by
\beq
    |\Psi_a\ket^\ft = \sum_b L^{d_b-d_a} M_a^b(mL)
    |\tilde\Psi_b\ket^\ft~.
\eeq
Define the operator $U\in\End(\ft)$ generating this mixing matrix:
\beq
    |\Psi_a\ket^\ft = U |\tilde\Psi_a\ket^\ft~.
\eeq
From the arguments above, it can be shown that
\beq
    U = \exp\lt[\int_{-\infty}^\infty d\theta
    \frc{A_-(\theta)A_+(\theta)}{1+e^{LE_\theta}}\rt]~.
\eeq

\sect{Riemann-Hilbert problem for finite-temperature form factors}

\label{sectRiemann}

\ssect{Non-interacting fields}

In Section \ref{sectftffunint}, we described finite-temperature
form factors of non-interacting fields. They are all given by
entire functions of the rapidity variables $\theta_i$'s, as are
zero-temperature form factors of non-interacting fields; the main
non-trivial phenomenon is that of mixing between a field and its
ascendants.

In fact one can verify that the linear space of functions $f^\Psi$
of the variables $\theta_i,\,i=1,\ldots,k$ with the properties:
\begin{enumerate}
\item $f^\Psi$ acquires a sign under exchange of any two of the
rapidity variables;
\item $f^\Psi$ has the quasi-periodicity property
(\ref{qpunint});
\item $f^\Psi$ is an entire functions of all its variables;
\end{enumerate}
reproduces the linear space of $k$-particle finite-temperature
form factors $f^\Psi_{+,\ldots,+}(\theta_1,\ldots,\theta_k;L)$ of
non-interacting operators $\Psi$. This constitutes a very simple
Riemann-Hilbert problem. Of course, the solution to this
Riemann-Hilbert problem also reproduces the set of
zero-temperature form factors; in order to disentangle the
finite-temperature form factors, one needs the additional
information about the mixing matrix.

\ssect{Riemann-Hilbert problem associated to right-twist
operators} \label{RHtwist}

As recalled in Appendix \ref{appTwist}, there are two types of
operators associated to each twist field: those with branch cut on
their right (``right-twist operators''), and those with branch cut
on their left (``left-twist operators''). As explained in Section
\ref{sectdefftff}, with an appropriate choice of the direction of
the branch cuts, one can obtain two-point functions of twist
fields in the NS sector and in the R sector. In this subsection,
we will consider right-twist operators only.

One can expect a description for the finite-temperature form
factors of twist fields (which are ``interacting'' fields) in the
same spirit as the one in the previous sub-section for
non-interacting fields.

Consider the function
\[
    f(\theta_1,\ldots,\theta_k;L) = f^{\Or_+}_{+,\ldots,+}(\theta_1,\ldots,\theta_k;L)
\]
where $\Or_+$ is the operator with branch cut on its right
representing a twist field: this can be the order field $\sigma_+$
or the disorder field $\mu_+$, or any of their conformal
descendants (that is, fields which reproduce conformal descendants
in the massless limit). Conformal descendants include space
derivatives, as well as other fields related to action of higher
conformal Virasoro modes on twist fields. A way of describing such
descendants is by taking the limit $x\to0$ of the finite part of
the OPE $\Psi(x)\sigma_+(0)$ or $\Psi(x)\mu_+(0)$, where $\Psi$ is
any bosonic non-interacting field.

The function $f$ solves the following Riemann-Hilbert problem:
\begin{enumerate}
\item Statistics of free particles: $f$ acquires a sign under exchange
of any two of the rapidity variables;
\item Quasi-periodicity:
    \[
    f(\theta_1,\ldots,\theta_j+2i\pi,\ldots,\theta_k;L)
     = - f(\theta_1,\ldots,\theta_j,\ldots,\theta_k;L)\,,\quad j=1,\ldots,k~;
    \]
\item Analytic structure: $f$ is analytic as function of all of its variables
$\theta_j,\,j=1,\ldots,k$ everywhere on the complex plane except
at simple poles. In the region ${\rm Im}(\theta_j) \in
[-i\pi,i\pi],\,j=1,\ldots,k$, its analytic structure is specified
as follows:
\begin{enumerate}
\item Thermal poles and zeroes: $f(\theta_1,\ldots,\theta_k;L)$ has poles at
    \[
    \theta_j=\alpha_n - \frc{i\pi}2 ,\;n\in\Z\,,\quad j=1,\ldots,k
    \]
    and has zeroes at
    \[
    \theta_j = \alpha_n-\frc{i\pi}2,\; n\in\Z+\frc12,\,\quad j=1,\ldots,k~,
    \]
\item Kinematical
    poles: $f(\theta_1,\ldots,\theta_k;L)$ has poles, as a function of $\theta_k$, at
    $\theta_{j}\pm i\pi,\,j=1,\ldots,k-1$ with residues given by \[
    f(\theta_1,\ldots,\theta_k;L)
    \sim \pm\frc{(-1)^{k-j}}\pi
    \frc{1+e^{-LE_{\theta_{j}}}}{1-e^{-LE_{\theta_{j}}}}
    \frc{f(\theta_1,\ldots,\h\theta_j,\ldots,\theta_{k-1};L)}{\theta_k-\theta_{j}\mp
    i\pi}~.
    \]
\end{enumerate}
\end{enumerate}
In order to have other finite-temperature form factors than those
with all positive charges, one more relation needs to be used. We
have:
\begin{itemize}
\item[4.] Crossing symmetry:
\[
    f_{\ep_1,\ldots,\ep_j,\ldots,\ep_k}^{\Or_+}(\theta_1,\ldots,\theta_j+i\pi,\ldots,\theta_k;L)
    = i
    f_{\ep_1,\ldots,-\ep_j,\ldots,\ep_k}^{\Or_+}(\theta_1,\ldots,\theta_j,\ldots,\theta_k;L)~.
\]
\end{itemize}
The same relation is also valid for non-interacting fields; recall
(\ref{crossingunint}). The name ``crossing symmetry'' is inspired
by the zero-temperature case. To make it more obvious, define the
functions
\beq\label{ftme}
    f(\theta_1',\ldots,\theta_l'|\theta_1,\ldots,\theta_k;L) =
    {\ }_{+,\ldots,+}\hspace{-4mm}{\ }^\ft
    \bra\theta_1',\ldots,\theta_l'|
    \Or_+^\ft(0)|\theta_1,\ldots,\theta_k\ket_{+,\ldots,+}^\ft~.
\eeq
These are in fact distributions, and can be decomposed in terms
supported at separated rapidities
$\theta_i'\neq\theta_j,\,\forall\,i,j$, and terms supported at
colliding rapidities, $\theta_i'=\theta_j$ for some $i$ and $j$.
We will denote the former by
$f^{sep.}(\theta_1',\ldots,\theta_l'|\theta_1,\ldots,\theta_k;L)$,
and the latter by
$f^{coll.}(\theta_1',\ldots,\theta_l'|\theta_1,\ldots,\theta_k;L)$.
Under integration over rapidity variables, the former gives
principal value integrals. Recalling the property
(\ref{invcharge}), we have
\[
    f^{sep.}(\theta_1',\ldots,\theta_l'|\theta_1,\ldots,\theta_k;L)
    =
    f^{\Or_+}_{+,\ldots,+,-,\ldots,-}(\theta_1,\ldots,\theta_k,\theta_l',\ldots,\theta_1';L)
\]
for
$(\theta_i'\neq\theta_j\;\forall\;i\in\{1,\ldots,l\},\,j\in\{1,\ldots,k\})$,
where on the right-hand side, there are $k$ positive charges
$(+)$, and $l$ negative charges $(-)$. Analytically extending from
its support the distribution $f^{sep.}$ to a function of complex
rapidity variables, crossing symmetry can then be written
\[\ba{l}\ds
    f^{sep.}(\theta_1',\ldots,\theta_l'|\theta_1,\ldots,\theta_k+i\pi;L)
    = i
    f^{sep.}(\theta_1',\ldots,\theta_l',\theta_k|\theta_1,\ldots,\theta_{k-1};L)
    ~,
    \z
    f^{sep.}(\theta_1',\ldots,\theta_l'+i\pi|\theta_1,\ldots,\theta_k;L)
    = i
    f^{sep.}(\theta_1',\ldots,\theta_{l-1}'|\theta_1,\ldots,\theta_k,\theta_l';L)
    ~,\ea
\]
which justifies its name.

It is worth mentioning that the distributive terms corresponding
to colliding rapidities satisfy a set of recursive equations:
\begin{itemize}
\item[5.] Colliding part of matrix elements:
\[\ba{l}
    f^{coll.}(\theta_1',\ldots,\theta_l'|\theta_1,\ldots,\theta_k;L)
    = \z \qquad
    \sum_{i=1}^l\sum_{j=1}^k (-1)^{l+k-i-j}\frc{1+e^{-LE_{\theta_j}}}{1-e^{LE_{\theta_j}}}
    \,\delta(\theta_i'-\theta_j) f(\theta_1',\ldots,\h\theta_i',\ldots,\theta_l'|
    \theta_1,\ldots,\h\theta_j,\ldots,\theta_k;L)~.
    \ea
\]
\end{itemize}
Note that the colliding part vanishes in the limit of zero
temperature, $L\to\infty$. Finally, it is instructive to re-write
the distribution
$f(\theta_1',\ldots,\theta_l'|\theta_1,\ldots,\theta_k;L)$ as an
analytical function with slightly shifted rapidities, plus a
distribution, using the relation
\beq
    \frc1{\theta-i0^+} = i\pi\delta(\theta) +
    \prin\lt(\frc1\theta\rt)
\eeq
where $\prin$ means that we must take the principal value integral
under integration. Defining the disconnected part
$f^{disconn.}(\theta_1',\ldots,\theta_l'|\theta_1,\ldots,\theta_k;L)$
of the matrix element (\ref{ftme}) as
\beq
    f(\theta_1',\ldots,\theta_l'|\theta_1,\ldots,\theta_k;L)
    = f^{sep.}(\theta_1'-i0^+,\ldots,\theta_l'-i0^+|\theta_1,\ldots,\theta_k;L)
    +
    f^{disconn.}(\theta_1',\ldots,\theta_l'|\theta_1,\ldots,\theta_k;L)
\eeq
where again we analytically extend from its support the
distribution $f^{sep.}$ to a function of complex rapidity
variables, we find that the disconnected part satisfies the
recursion relations
\[\ba{l}
    f^{disconn.}(\theta_1',\ldots,\theta_l'|\theta_1,\ldots,\theta_k;L)
    = \z \qquad
    \sum_{i=1}^l\sum_{j=1}^k (-1)^{l+k-i-j}\,(1+e^{-LE_{\theta_j}})
    \,\delta(\theta_i'-\theta_j) f(\theta_1',\ldots,\h\theta_i',\ldots,\theta_l'|
    \theta_1,\ldots,\h\theta_j,\ldots,\theta_k;L)~.
    \ea
\]
Note that the factor $(1+e^{-LE_{\theta_j}})
\,\delta(\theta_i'-\theta_j)$ appearing inside the double sum is
just the overlap ${\ }_{+}\hspace{-4mm}{\ }^\ft
    \bra\theta_i'|\theta_j\ket_{+}^\ft$, so that the equation
above can be naturally represented as a sum of disconnected
diagrams.

We will derive below all of the three points in the
Riemann-Hilbert problem above, as well as crossing symmetry and
the recursive formula for the colliding part of matrix elements.

\ssect{Mixing for twist fields}

\label{mixingtwist}

It is natural to suppose that, like in the case of non-interacting
fields, some ``mixing'' occurs between a twist field $\Or$ and its
ascendants -- fields of lower dimension and with the same locality
index and statistics as those of $\Or$ -- in calculating the
finite-temperature form factors. We describe here a conjecture for
the way mixing should occur. We will use right-twist operators for
this description, but we expect that the same mixing should occur
amongst left-twist operators.

First note that the fact that finite-temperature form factors have
exponential asymptotic behavior, when all rapidity variables are
sent to positive or negative infinity simultaneously, can be seen
as following from the requirement of convergence of the form
factor expansion (\ref{ftffexp}) in the region $0<\tau<L$.
Consider a solution $\t{f}_{\t\Or_+}$ to the Riemann-Hilbert
problem above (Points 1, 2 and 3), with the property that
\[
    \t{f}_{\t\Or_+}(\theta_1+\beta,\ldots,\theta_k+\beta;L) \sim
    \bra\vac|\t\Or_+(0)|\theta_1,\ldots,\theta_k\ket\;
    e^{s\beta} \quad\mbox{as}\quad \beta\to\pm\infty
\]
for some twist operators $\t\Or_+$ of spins $s$. It will be clear
in the next section that such solutions exist. Consider a mixing
matrix for twist fields whose elements $M_{\Or_+}^{\t\Or_+}(mL)$
are parametrized by two operators $\Or_+$ and $\t\Or_+$. This
mixing matrix is characteristic of the theory, and its elements
are non-zero only for operators $\t\Or_+$ that are ascendants of
$\Or_+$. The finite-temperature form factors of $\Or_+$ are the
following linear combinations:
\beq
    f(\theta_1,\ldots,\theta_k;L) = \sum_{\t\Or_+}
    M_{\Or_+}^{\t\Or_+}(m L)
    \t{f}_{\t\Or_+}(\theta_1,\ldots,\theta_k;L)~.
\eeq
We do not yet have a full derivation of this, neither a
description of the mixing matrix, but it seems a natural
assumption.

\ssect{Derivation of the Riemann-Hilbert problem associated to
right-twist operators}

\label{RHderivation}

{\bf Point 1.}

The first point of the Riemann-Hilbert problem of sub-section
\ref{RHtwist} is a direct consequence of the definition of the
finite-temperature form factors and of the canonical
anti-commutation relations of the free fermionic modes.

{\bf Point 3a.}

We now derive the position of the residues and zeroes stated in
Point 3a. We first concentrate on one-particle finite-temperature
form factors. Consider the two-point functions
\beq\label{g}
    g(x,\tau) = {\ }^\ft\bra\vac|\Or_+^\ft(0,0)\psi^\ft(x,\tau)|\vac\ket^\ft~.
\eeq
and
\beq\label{gt}
    \t{g}(x,\tau) = {\ }^\ft\bra\vac|\psi^\ft(x,\tau)\Or_+^\ft(0,0)|\vac\ket^\ft~.
\eeq
Their finite-temperature form factor expansions are, from
(\ref{ftfffermions}),
\beq\label{ftffexpg}
    g(x,\tau) =  \frc12\sqrt{\frc{m}{\pi}}\,\int d\theta\,
    e^{\theta/2} \lt(
    \frc{f^{\Or_+}_+(\theta;L)}{1+e^{-LE_\theta}} \,e^{-ip_\theta x + E_\theta\tau} +
    \frc{f^{\Or_+}_-(\theta;L)}{1+e^{LE_\theta}} \,e^{ip_\theta x - E_\theta\tau}
    \rt)
\eeq
and
\beq\label{ftffexpgt}
    \t{g}(x,\tau) =  \frc12\sqrt{\frc{m}{\pi}}\,\int d\theta\,
    e^{\theta/2} \lt(
    \frc{f^{\Or_+}_-(\theta;L)}{1+e^{-LE_\theta}} \,e^{ip_\theta x - E_\theta\tau} +
    \frc{f^{\Or_+}_+(\theta;L)}{1+e^{LE_\theta}} \,e^{-ip_\theta x + E_\theta\tau}
    \rt)~.
\eeq
The one-particle form factors are essentially fixed, up to
normalization, by the requirements:
\begin{itemize}
\item convergence of the form factor expansion of $g(x,\tau)$ (\ref{ftffexpg})
in the region $-L<\tau<0$;
\item quasi-periodicity of the analytic continuation in $\tau$: $g(x,\tau+L) = -
g(x,\tau)$ if $x<0$ and $g(x,\tau+L) = g(x,\tau)$ if $x>0$.
\end{itemize}
For $\t{g}(x,\tau)$, the first requirement is replaced by a
convergence in $0<\tau<L$, and the second requirement holds
unchanged; they also essentially fix the one-particle form
factors. The first requirement implies that the form factors
$f^{\Or_+}_\pm(\theta;L)$ must go as $\propto e^{p\theta}$ for
some number $p$ as $\theta\to\infty$, and as $\propto e^{q\theta}$
for some number $q$ as $\theta\to-\infty$. The second specifies
the positions of poles and zeros.

The second requirement will be satisfied if we can deform the
contour in $\theta$-space and take poles at appropriate values.
With the generalization to multi-particle finite-temperature form
factors in mind, we will consider these deformations in
$g(x,\tau)$ for $x<0$ only and in $\t{g}(x,\tau)$ for $x>0$ only.
For $x<0$, in (\ref{ftffexpg}), we can shift the $\theta$-contour
upwards for the term containing $e^{-ip_\theta x}$ and downwards
for the term containing $e^{ip_\theta x}$. Consider the functions
\[
    g_+(\theta) =
    \frc{f_+^{\Or_+}(\theta;L)}{1+e^{-LE_\theta}}~,\qquad
    g_-(\theta) = \frc{f_-^{\Or_+}(\theta;L)}{1+e^{LE_\theta}}~.
\]
The anti-periodicity condition at $x<0$ will be satisfied if: the
function $g_+(\theta)$ has poles at $\theta = \alpha_n+i\pi/2$
(recall (\ref{defalphan})) for $n\in\Z+\frc12$; the function
$g_-(\theta)$ has poles at $\theta = \alpha_n-i\pi/2$ for
$n\in\Z+\frc12$; and $e^{i\pi/4}g_+(\theta'+i\pi/2) +
e^{-i\pi/4}g_-(\theta'-i\pi/2) = 0$ for all $\theta'$ real (except
at the positions of the poles). Then, by deforming the contours,
we get the sum of $2\pi i$ times the residues of the poles of
$g_+(\theta)$ (equivalently, $-2\pi i$ times the residues of the
poles of $g_-(\theta)$) and the resulting function is
anti-periodic in $\tau$. For $x>0$, consider (\ref{ftffexpgt}).
Define
\[
    \t{g}_+(\theta) =
    \frc{f_+^{\Or_+}(\theta;L)}{1+e^{LE_\theta}}~,\qquad
    \t{g}_-(\theta) = \frc{f_-^{\Or_+}(\theta;L)}{1+e^{-LE_\theta}}~.
\]
With similar arguments for the function $\t{g}(x,\tau)$ to be
periodic: the function $\t{g}_+(\theta)$ has poles at $\theta =
\alpha_n-i\pi/2$ for $n\in\Z$; the function $g_-(\theta)$ has
poles at $\theta = \alpha_n+i\pi/2$ for $n\in\Z$; and
$e^{-i\pi/4}g_+(\theta'-i\pi/2) + e^{i\pi/4}g_-(\theta'+i\pi/2) =
0$ for all $\theta'$ real (except at the positions of the poles).
There must be no other poles in ${\rm Im}(\theta) \in
[-\pi/2,\pi/2]$ for the four functions $g_\pm(\theta)$ and
$\t{g}_\pm(\theta)$ (note that $g_\pm(\theta)/\t{g}_\pm(\theta)$
are entire functions of $\theta$). Below we will argue that there
are no other poles in the wider range ${\rm Im}(\theta) \in
[-\pi,\pi]$. Assuming further no other types of singularities than
simple poles, we conclude that $g_\pm(\theta)$ and
$\t{g}_\pm(\theta)$ must have no other singularities than those
mentioned above in the region ${\rm Im}(\theta) \in [-\pi,\pi]$.

Recall, from the general relation (\ref{invcharge}), that
$f_-^{\Or_+}(\theta;L) = \lt(f_+^{\Or_+}(\theta;L)\rt)^\dag$. For
the function $f_+^{\Or_+}(\theta;L)$, this gives the following
conditions:
\begin{itemize}
\item  $f_+^{\Or_+}(\theta;L)$ has poles at $\theta=\alpha_n
    - \frc{i\pi}2 ,\;n\in\Z$ and has zeroes at $\theta =
    \alpha_n-\frc{i\pi}2,\; n\in\Z+\frc12$;
\item $f_+^{\Or_+}(\theta;L)$ does not have poles for ${\rm
    Im}(\theta) \in [-\pi,\pi]$ except for those mentioned
    above;
\item ${\rm Re}\lt(e^{\frc{-i \pi}4}f_+^{\Or_+}(\theta'-i\pi/2;L)\rt) =
    0~,\quad
    {\rm Re}\lt(e^{\frc{i \pi}4}f_+^{\Or_+}(\theta'+i\pi/2;L)\rt) =
    0\quad$
    for $\theta'\in\Re$.
\end{itemize}
The first and second points are the one-particle case of Point 3a.
We will verify, in explicit calculations of finite-temperature
form factors below, that the last point is in fact a consequence
of the full Riemann-Hilbert problem of sub-section (\ref{RHtwist})
along with (\ref{invcharge}).

The generalization to multi-particle finite-temperature form
factors
$f_{\ep_1,\ldots,\ep_k}^{\Or_+}(\theta_1,\ldots,\theta_k;L)$ goes
along the same lines. We need to calculate the multi-point
function with $k$ insertions of the fermion field $\psi$ at $k$
different points. We need to consider two cases: one with these
insertions on the right of $\Or_+$, and one with these insertions
on its left. We only have to assume that there are no poles in
$|{\rm Im}(\theta_i-\theta_j)|\leq \pi/2$ if the two rapidities
$\theta_i$ and $\theta_j$ are associated to the same charge, and
in $|{\rm Im}(\theta_i-\theta_j)|\leq \pi$, except possibly at
$\theta_i=\theta_j$, if the two rapidities are associated to
opposite charges. The poles at $\theta_i=\theta_j$ (below, we will
show that they are present) are taken care of by the prescription
stated at the end of Section \ref{sectdefftff}. This prescription
does not affect our arguments, since we were careful to shift
rapidity contours in directions making the two-point function
convergent at $\tau=0$ with positive difference between the
positions of the first and of the second operator. Then, repeating
the arguments, we find Point 3a.

{\bf Points 2 and 4.}

Point 2 is evidently a consequence of Point 4. Hence we will show
the latter.

We concentrate again, first, on the one-particle form factors.
Consider the function $g(x,\tau)$ (\ref{g}) and its form factor
expansion (\ref{ftffexpg}). Consider $x<0$, and shift the contour
in the first term as $\theta\mapsto\theta+i\pi/2$ and in the
second term as $\theta\mapsto\theta-i\pi/2$. This gives, as
explained above, a sum of residues. The same sum of residues can
be obtained, up to an overall minus sign, by taking the
$\theta$-contour slightly above the line of imaginary part $\pi/2$
in the first term, and slightly below the line of imaginary part
$-\pi/2$ in the second term. This gives again a representation of
the two-point function valid in the region $-L<\tau<0$, $x<0$.
From this representation, we can shift the $\theta$-contours all
the way to the line of imaginary part $\pi$ in the first term, and
to the line of imaginary part $-\pi$ in the second term, taking
residues of poles, if any. We obtain a representation of exactly
the same form as (\ref{ftffexpg}), valid in the same regions of
$x$ and $\tau$. Since a representation of this form is unique,
each coefficient of the exponentials should be the same; in
particular, there should be no poles in the region $\pi/2<{\rm
Im}(\theta)\leq\pi$ for the first term and $-\pi<{\rm
Im}(\theta)\leq-\pi/2$ for the second. The same argument can be
applied to the representation (\ref{ftffexpgt}) of the function
$\t{g}(x,\tau)$ in the region $x>0$. The fact that the
representation is unique, taking into account the prefactor
$e^{\theta/2}$ in (\ref{ftffexpg}) and (\ref{ftffexpgt}) and the
minus sign occurring when going beyond the line of poles at
imaginary parts $\pm \pi/2$, gives
\[
    f^{\Or_+}_\pm(\theta+i\pi;L) = if^{\Or_+}_\mp(\theta;L)~,
\]
which is the one-particle case of Point 4.

The generalization to multi-particle form factors goes along the
same lines. The prescription stated at the end of Section
\ref{sectdefftff}, for the case of poles appearing at colliding
rapidity variables, stays invariant under shifting by $\pm i\pi$
the rapidity integration lines in the argument above.

{\bf Points 3b and 5.}

We now prove Points 3b and 5. We first concentrate on the
two-particle case. Consider the distribution
\beq
    h(\theta_2|\theta_1) \equiv \braL a(\theta_2)
    \Or_+(0)
    a^\dag(\theta_1)\ketL
    = \frc{f(\theta_2|\theta_1;L)}{(1+e^{-LE_{\theta_1}})(1+e^{-LE_{\theta_2}})}
\eeq
where on the right-hand side we use the definition (\ref{ftme}).
Expanding the modes in fermion fields using (\ref{modespsi}), we
have
\beqa
    h(\theta_2|\theta_1) &=&
    \frc{m}{4\pi} \int dx_2dx_1\,e^{-ip_2x_2+ip_1x_1} \times
    \n&&\qquad\times\;
        \Bigg[e^{\frc{\theta_2+\theta_1}2} \braL
        \psi(x_2)\Or_+(0)\psi(x_1)\ketL -i e^{\frc{\theta_2-\theta_1}2} \braL
        \psi(x_2)\Or_+(0)\bar\psi(x_1)\ketL \n&&\qquad\quad
        +i e^{-\frc{\theta_2-\theta_1}2} \braL\bar\psi(x_2)\Or_+(0)\psi(x_1)\ketL
        +e^{-\frc{\theta_2+\theta_1}2} \braL\bar\psi(x_2)\Or_+(0)\bar\psi(x_1)\ketL
        \Bigg]~.\no
\eeqa
Consider the first term inside the bracket. We want the main
behavior of the expression around $\theta_2=\theta_1$. It is
obtained by taking $x_2$ and $x_1$ very large, both with the same
sign. In this limit, the correlation functions factorize, so the
main contribution of the first term is contained into
\beq
    \frc{m}{4\pi} \int dx_2dx_1\,e^{-ip_2x_2+ip_1x_1}\,
    e^{\frc{\theta_2+\theta_1}2} {\rm sign}(-x_1)\braL
    \psi(x_2)\psi(x_1)\ketL^{{\rm sign}(x_1)} \braL\Or_+(0)\ketL
\eeq
where the correlation function of fermion fields is taken with
{\em periodic} conditions around the cylinder if $x_1>0$, and
anti-periodic conditions if $x_1<0$:
\[
    \braL \cdots\ketL^\pm \quad:\quad \mbox{trace with periodic $(+)$ /
    anti-periodic $(-)$ conditions on the fermion fields}~.
\]
The factor ${\rm sign}(-x_1)$ arises because we must put the
operator $\psi(x_2)$ at a slightly positive Euclidean time
$\tau_2=0^+$ and $\psi(x_1)$ at a slightly negative Euclidean time
$\tau_1=0^-$ in the initial correlation function, and we must take
$\psi(x_1)$ through the branch cut produced by $\Or_+(0)$, if
$x_1>0$, before we can factorize. Changing variables, this is
\beq
    \frc{m}{4\pi} \int dx_2dx_1\,e^{-ip_2x_2+i(p_1-p_2)x_1}\,
    e^{\frc{\theta_2+\theta_1}2} {\rm sign}(-x_1)\braL
    \psi(x_2)\psi(0)\ketL^{{\rm sign}(x_1)} \braL\Or_+(0)\ketL~.
\eeq
The integral over $x_1$ is a sum of two distributions, one
supported on $p_2-p_1\neq0$, one supported at $p_2-p_1=0$. They
can be evaluated by using the distributional equations
\beq
    \int dx\, e^{ipx} {\rm sign}(x) = 2i \,\prin\lt(\frc1p\rt)~,\quad
    \int dx\, e^{ipx} = 2\pi \delta(p)
\eeq
where $\prin$ means principal value. The part supported on
$p_2-p_1\neq0$ is
\beq
    \frc{im}{4\pi} \,\prin\lt(\frc{1}{p_2-p_1}\rt)
    \int dx_2\,e^{-ip_2x_2}\,
    e^{\frc{\theta_2+\theta_1}2} \lt[ \braL
    \psi(x_2)\psi(0)\ketL^{+}
    +
    \braL \psi(x_2)\psi(0)\ketL^{-}
    \rt]\braL\Or_+(0)\ketL
\eeq
whereas the part supported at $p_2-p_1=0$ is
\beq
    \frc{m}{4} \,\delta(p_2-p_1)\int dx_2\,e^{-ip_2x_2}\,
    e^{\frc{\theta_2+\theta_1}2} \lt[\braL
    \psi(x_2)\psi(0)\ketL^{-} - \braL
    \psi(x_2)\psi(0)\ketL^{+}\rt] \braL\Or_+(0)\ketL~.
\eeq
Putting all terms together, the former gives
\beq
    h^{sep.}(\theta_2|\theta_1) \sim \frc{im}{4\pi}
    \,\prin\lt(\frc{1}{\theta_2-\theta_1}\rt)\frc{1}{E_{\theta_1}}
    \int dx_2\,e^{-ip_1x_2}\,\lt[ \braL
    \Psi(\theta_1|x_2)\ketL^{+}
    +
    \braL \Psi(\theta_1|x_2)\ketL^{-}
    \rt]\braL\Or_+(0)\ketL
\eeq
and the latter gives
\beq
    h^{coll.}(\theta_2|\theta_1) = \frc{m}{4}
    \delta(\theta_2-\theta_1)\frc{1}{E_{\theta_1}}
    \int dx_2\,e^{-ip_1x_2}\,\lt[ \braL
    \Psi(\theta_1|x_2)\ketL^{-}
    -
    \braL \Psi(\theta_1|x_2)\ketL^{+}
    \rt]\braL\Or_+(0)\ketL
\eeq
where
\[
    \Psi(\theta_1|x_2) = e^{\theta_1}\psi(x_2)\psi(0) - i
    \psi(x_2)\b\psi(0)+i\b\psi(x_2)\psi(0)+e^{-\theta_1}\b\psi(x_2)\b\psi(0)~.
\]
In order to evaluate the integral over $x_2$, consider the traces
\beqa
    \braL a(\theta_2) a^\dag(\theta_1) \ketL^- &=&
    \frc{\delta(\theta_2-\theta_1)}{1+e^{-LE_{\theta_1}}} \n
    \braL a(\theta_2)a^\dag(\theta_1)\ketL^+ &=&
    \frc{\delta(\theta_2-\theta_1)}{1-e^{-LE_{\theta_1}}}~. \no
\eeqa
A derivation similar to the one above but now applied to these
objects gives
\beq
    \frc{m}{2}
    \delta(p_2-p_1)
    \int dx_2\,e^{-ip_1x_2}\,\braL
    \Psi(\theta_1|x_2)\ketL^{\pm} = \frc{\delta(\theta_2-\theta_1)}{1\mp
    e^{-LE_{\theta_1}}}~.
\eeq
Hence,
\beqa
    h^{sep.}(\theta_2|\theta_1) &\sim& \frc{i}{2\pi}
    \,\prin\lt(\frc{1}{\theta_2-\theta_1}\rt)\lt[
    \frc1{1-e^{-LE_{\theta_1}}} +
    \frc1{1+e^{-LE_{\theta_1}}}\rt]\;\braL\Or_+(0)\ketL
    \n
    &=& \frc{i}{\pi}
    \,\prin\lt(\frc{1}{\theta_2-\theta_1}\rt)
    \frc1{(1-e^{-LE_{\theta_1}})(1+e^{-LE_{\theta_1}})}\;\braL\Or_+(0)\ketL
\eeqa
from which we have
\beq
    f^{sep.}(\theta_2|\theta_1;L) \sim
    \frc{i}{\pi}\frc{1+e^{-LE_{\theta_1}}}{1-e^{-LE_{\theta_1}}}
    \,\prin\lt(\frc1{\theta_2-\theta_1}\rt)
    \;\braL\Or_+(0)\ketL~.
\eeq
Combined with crossing symmetry, this proves Point 3b for the case
$k=2,\,j=1$. On the other hand, the delta-function part is given
by
\beqa
    h^{coll.}(\theta_2|\theta_1) &=&
    \frc12\delta(\theta_2-\theta_1)\lt[
    \frc1{1+e^{-LE_{\theta_1}}} -
    \frc1{1-e^{-LE_{\theta_1}}}\rt]\;\braL\Or_+(0)\ketL
    \n
    &=&
    \frc{\delta(\theta_2-\theta_1)}{
    (1-e^{LE_{\theta_1}})(1+e^{-LE_{\theta_1}})}\;\braL\Or_+(0)\ketL
\eeqa
which shows Point 5 for $l=k=1,\,i=j=1$.

A similar argument holds for $k>2$, with extra minus signs coming
from the odd statistics of the fermion fields, of their modes, and
of the twist operator if it has non-zero form factors with odd
particle numbers only.

\ssect{Riemann-Hilbert problem associated to left-twist operators}

\label{RHltwist}

Similar arguments can be used to analyze finite-temperature form
factors of twist fields with branch cut on their left. We only
state here the corresponding Riemann-Hilbert problem.

Consider the function
\[
    f(\theta_1,\ldots,\theta_k;L) = f^{\Or_-}_{+,\ldots,+}(\theta_1,\ldots,\theta_k;L)
\]
where $\Or_-$ is the operator with branch cut on its left
representing a twist field.

The function $f$ solves the following Riemann-Hilbert problem:
\begin{enumerate}
\item Statistics of free particles: $f$ acquires a sign under exchange
of any two of the rapidity variables;
\item Quasi-periodicity:
    \[
    f(\theta_1,\ldots,\theta_j+2i\pi,\ldots,\theta_k;L)
     = - f(\theta_1,\ldots,\theta_j,\ldots,\theta_k;L)\,,\quad j=1,\ldots,k~;
    \]
\item Analytic structure: $f$ is analytic as function of all of its variables
$\theta_j,\,j=1,\ldots,k$ everywhere on the complex plane except
at simple poles. In the region ${\rm Im}(\theta_j) \in
[-i\pi,i\pi],\,j=1,\ldots,k$, its analytic structure is specified
as follows:
\begin{enumerate}
\item Thermal poles and zeroes: $f(\theta_1,\ldots,\theta_k;L)$ has poles at
    \[
    \theta_j=\alpha_n + \frc{i\pi}2 ,\;n\in\Z\,,\quad j=1,\ldots,k
    \]
    and has zeroes at
    \[
    \theta_j = \alpha_n+\frc{i\pi}2,\; n\in\Z+\frc12,\,\quad j=1,\ldots,k~,
    \]
\item Kinematical poles: $f(\theta_1,\ldots,\theta_k;L)$ has poles,
    as a function of $\theta_k$, at
    $\theta_{j}\pm i\pi,\,j=1,\ldots,k-1$ with residues given by \[
    f(\theta_1,\ldots,\theta_k;L)
    \sim \mp\frc{(-1)^{k-j}}\pi
    \frc{1+e^{-LE_{\theta_{j}}}}{1-e^{-LE_{\theta_{j}}}}
    \frc{f(\theta_1,\ldots,\h\theta_j,\ldots,\theta_{k-1};L)}{\theta_k-\theta_{j}\mp
    i\pi}~.
    \]
\end{enumerate}
\end{enumerate}
Again, in order to have other finite-temperature form factors than
those with all positive charges, one more relation needs be used.
We have:
\begin{itemize}
\item[4.] Crossing symmetry:
\[
    f_{\ep_1,\ldots,\ep_j,\ldots,\ep_k}^{\Or_-}(\theta_1,\ldots,\theta_j+i\pi,\ldots,\theta_k;L)
    = i
    f_{\ep_1,\ldots,-\ep_j,\ldots,\ep_k}^{\Or_-}(\theta_1,\ldots,\theta_j,\ldots,\theta_k;L)~.
\]
\end{itemize}
Moreover, matrix elements\[
    f(\theta_1',\ldots,\theta_l'|\theta_1,\ldots,\theta_k;L) =
    {\ }_{+,\ldots,+}\hspace{-4mm}{\ }^\ft
    \bra\theta_1',\ldots,\theta_l'|
    \Or_-^\ft(0)|\theta_1,\ldots,\theta_k\ket_{+,\ldots,+}^\ft
\]
can again be decomposed in terms supported at separated rapidities
$\theta_i'\neq\theta_j,\,\forall\,i,j$ (which give principal value
integrals under integration), and terms supported at colliding
rapidities, $\theta_i'=\theta_j$ for some $i$ and $j$, denoted
respectively by
$f^{sep.}(\theta_1',\ldots,\theta_l'|\theta_1,\ldots,\theta_k;L)$
and
$f^{coll.}(\theta_1',\ldots,\theta_l'|\theta_1,\ldots,\theta_k;L)$.
Recalling the property (\ref{invcharge}), we have
\[
    f^{sep.}(\theta_1',\ldots,\theta_l'|\theta_1,\ldots,\theta_k;L)
    =
    f^{\Or_-}_{+,\ldots,+,-,\ldots,-}(\theta_1,\ldots,\theta_k,\theta_l',\ldots,\theta_1';L)
\]
for
$(\theta_i'\neq\theta_j\;\forall\;i\in\{1,\ldots,l\},\,j\in\{1,\ldots,k\})$,
where on the right-hand side, there are $k$ positive charges
$(+)$, and $l$ negative charges $(-)$. The distributive terms
corresponding to colliding rapidities satisfy the same set of
recursive equations as in the case of right-twist operators:
\begin{itemize}
\item[5.] Colliding part of matrix elements:
\[\ba{l}
    f^{coll.}(\theta_1',\ldots,\theta_l'|\theta_1,\ldots,\theta_k;L)
    = \z \qquad
    \sum_{i=1}^l\sum_{j=1}^k (-1)^{l+k-i-j}\frc{1+e^{-LE_{\theta_j}}}{1-e^{LE_{\theta_j}}}
    \,\delta(\theta_i'-\theta_j) f(\theta_1',\ldots,\h\theta_i',\ldots,\theta_l'|
    \theta_1,\ldots,\h\theta_j,\ldots,\theta_k;L)~.
    \ea
\]
\end{itemize}
Finally, we can again re-write the distribution
$f(\theta_1',\ldots,\theta_l'|\theta_1,\ldots,\theta_k;L)$ as an
analytical function with slightly shifted rapidities, plus a
distribution, using this time the relation
\beq
    \frc1{\theta+i0^+} = -i\pi\delta(\theta) +
    \prin\lt(\frc1\theta\rt)~.
\eeq
Defining the disconnected part
$f^{disconn.}(\theta_1',\ldots,\theta_l'|\theta_1,\ldots,\theta_k;L)$
of the matrix element (\ref{ftme}) as
\beq
    f(\theta_1',\ldots,\theta_l'|\theta_1,\ldots,\theta_k;L)
    = f^{sep.}(\theta_1'+i0^+,\ldots,\theta_l'+i0^+|\theta_1,\ldots,\theta_k;L)
    +
    f^{disconn.}(\theta_1',\ldots,\theta_l'|\theta_1,\ldots,\theta_k;L)
\eeq
where again we analytically extend from its support the
distribution $f^{sep.}$ to a function of complex rapidity
variables, we find that the disconnected part satisfies the
recursion relations
\[\ba{l}
    f^{disconn.}(\theta_1',\ldots,\theta_l'|\theta_1,\ldots,\theta_k;L)
    = \z \qquad
    \sum_{i=1}^l\sum_{j=1}^k (-1)^{l+k-i-j}\,(1+e^{-LE_{\theta_j}})
    \,\delta(\theta_i'-\theta_j) f(\theta_1',\ldots,\h\theta_i',\ldots,\theta_l'|
    \theta_1,\ldots,\h\theta_j,\ldots,\theta_k;L)~.
    \ea
\]

\sect{Finite-temperature form factors of twist fields}

\label{sectftfftwist}

For the order and disorder fields, $\sigma$ and $\mu$ (again, see
Appendix \ref{appTwist} for the definition of the associated
operators $\sigma_\pm$ and $\mu_\pm$), the solutions to the
Riemann-Hilbert problem of sub-section \ref{RHtwist} are
completely fixed (up to a normalization) by the asymptotic
behavior $\sim O(1)$ at $\theta\to\pm\infty$, since they are
primary fields of spin 0. Note that the method of computing
one-particle finite-temperature form factors by solving the
Riemann-Hilbert problem with this asymptotic behavior is very
similar to the method used by Fonseca and Zamolodchikov
\cite{FonsecaZ01} for calculating form factors on the circle.

For the one-particle finite-temperature form factor of the
disorder operator with a branch cut on its right, the solution is
\beq\label{1pftffmu+}
    f_\pm^{\mu_+}(\theta;L) = e^{\pm\frc{i\pi}4} C(L)\,
    \exp\lt[\mp\int_{-\infty\mp i0^+}^{\infty \mp i0^+} \frc{d\theta'}{2\pi i}
    \frc1{\sinh(\theta-\theta')}
    \ln\lt(\frc{1+e^{-LE_{\theta'}}}{1-e^{-LE_{\theta'}}}\rt)\rt]
\eeq
for some real constant $C(L)$. This is in agreement with the
Hermiticity of $\mu_+$, which gives $(f_\pm^{\mu_+}(\theta;L))^* =
f_\mp^{\mu_+}(\theta;L)$ for $\theta$ real. Using
\[
    \frc1{\sinh(\theta-(\theta'\pm i0^+))} = \pm i\pi \delta(\theta-\theta') +
    \prin\lt(\frc1{\sinh(\theta-\theta')}\rt)~.
\]
this can also be written
\beq
    f_\pm^{\mu_+}(\theta;L)
    = C(L) e^{\pm\frc{i\pi}4}
    \sqrt{\frc{1+e^{-LE_\theta}}{1-e^{-LE_\theta}}}\;
    \exp\lt[\mp\int_{-\infty}^\infty \frc{d\theta'}{2\pi i}
    \prin\lt(\frc{1}{\sinh(\theta-\theta')}\rt)
    \ln\lt(\frc{1+e^{-LE_{\theta'}}}{1-e^{-LE_{\theta'}}}\rt)\rt]~.
\eeq
That this is a solution can be checked by verifying the asymptotic
behavior $f_\pm^{\mu_+}(\theta;L) \sim e^{\pm\frc{i\pi}4}C(L)$ as
$|\theta|\to\infty$, and by verifying that the functions
$f_\pm^{\mu_+}(\theta;L)$ have poles and zeros at the proper
positions. Positions of poles and zeros are the values of $\theta$
such that when analytically continued from real values, a pole at
$\sinh(\theta-\theta')=0$ in the integrand of (\ref{1pftffmu+})
and one of the logarithmic branch points pinch the $\theta'$
contour of integration. The fact that these positions correspond
to poles and zeros can be deduced most easily from the functional
relation
\beq\label{fcteqn}
    f_\pm^{\mu_+}(\theta;L)f_\pm^{\mu_+}(\theta\pm i\pi;L) = \pm i C(L)^2
    \frc{1+e^{-LE_\theta}}{1-e^{-LE_\theta}}~.
\eeq
Note that this implies the quasi-periodicity property
\beq
    f_\pm^{\mu_+}(\theta+2i\pi;L) = -f_\pm^{\mu_+}(\theta;L)~.
\eeq
It is also easy to see that the crossing symmetry relation is
satisfied. Also, since $C(L)$ is real, one can check the validity
of the relation ${\rm Re}\lt(e^{\frc{-i
\pi}4}f_+^{\Or_+}(\theta'-i\pi/2;L)\rt) = 0~,\quad {\rm
Re}\lt(e^{\frc{i \pi}4}f_+^{\Or_+}(\theta'+i\pi/2;L)\rt) = 0\quad$
for $\theta'$ real; this relation was seen as a consequence of
general principles in the proof of Point 3a in sub-section
\ref{RHderivation}; it is now seen as a consequence of the
Riemann-Hilbert problem along with Hermiticity.

For the operator $\mu_-$ with a branch cut on its left, one can
check similarly that the function
\beq
    f_\pm^{\mu_-}(\theta;L) = f_\pm^{\mu_+}(\theta-i\pi;L) =
    -if_\mp^{\mu_+}(\theta;L)
\eeq
solves the Riemann-Hilbert problem of sub-section \ref{RHltwist}.
Explicitly,
\beq\label{1pftffmu-}
    f_\pm^{\mu_-}(\theta;L) = -ie^{\mp\frc{i\pi}4} C(L)\,
    \exp\lt[\pm\int_{-\infty\pm i0^+}^{\infty \pm i0^+} \frc{d\theta'}{2\pi i}
    \frc1{\sinh(\theta-\theta')}
    \ln\lt(\frc{1+e^{-LE_{\theta'}}}{1-e^{-LE_{\theta'}}}\rt)\rt]~.
\eeq
In particular, we observe that $(f_\pm^{\mu_-}(\theta;L))^* =
-f_\mp^{\mu_-}(\theta;L)$, which is in agreement with the
anti-Hermiticity of the operator $\mu_-$ (see Appendix
\ref{appTwist}). Note that we chose the same constant $C(L)$ as a
normalization for both $f_\pm^{\mu_-}$ and $f_\pm^{\mu_+}$. This
is not a consequence of the Riemann-Hilbert problem, but can be
checked by explicitly calculating the normalization. The
normalization is calculated in Appendix \ref{appnorm}, and is
given by
\beq\label{normC}
    C(L) = \frc{\braL \sigma \ketL}{\sqrt{2\pi}}
\eeq
where the average $\braL\sigma\ketL$ was calculated in
\cite{Sachdev96} (the average at zero-temperature (that is, $L\to\infty$) can
be found in \cite{McCoyWu}) and is given by
\[
    m^{\frc18}2^{\frc1{12}} e^{-\frc18} A^{\frc32} \exp\lt[ \frc{(mL)^2}2
    \int\int_{-\infty}^\infty \frc{d\theta_1d\theta_2}{(2\pi)^2}
    \frc{\sinh\theta_1\sinh\theta_2}{\sinh(mL\cosh\theta_1)\sinh(mL\cosh\theta_2)}
    \ln\lt|\lt(\coth\frc{\theta_1-\theta_2}2\rt)\rt|\rt]
\]
where $A$ is Glaisher's constant.

Multi-particle finite-temperature form factors can be easily
constructed from the well-known zero-temperature form factors
(first calculated in \cite{BergKW79}), by adjoining ``leg
factors'', which are just normalized one-particle
finite-temperature form factors:
\beq
    f^{\Or_+}_{+,\ldots,+}(\theta_1,\ldots,\theta_k;L)
    = i^{\lt[\frc{k}2\rt]}  \braL\sigma\ketL
    \lt(\prod_{j=1}^k
    \frc{f^{\mu_+}_+(\theta_j;L)}{\braL\sigma\ketL}
    \rt)
    \; \prod_{1\leq i<j\leq k}
    \tanh\lt(\frc{\theta_j-\theta_i}2\rt)
\eeq
where $\Or_+$ is $\sigma_+$ if $k$ is even, and $\mu_+$ if $k$ is
odd. The symbol $[k/2]$ equals the greatest integer smaller than
or equal to $k/2$. This satisfies the condition on thermal poles
and zeroes simply from the properties of the leg factors, and it
can be verified that this satisfies the quasi-periodicity
condition and the kinematical pole condition, Point 2 and Point 3b
of sub-section \ref{RHtwist}, respectively. Using crossing
symmetry, Point 4, it is a simple matter to obtain the formula for
other values of the charges:
\beq\label{ftfftwistp}
    f^{\Or_+}_{\ep_1,\ldots,\ep_k}(\theta_1,\ldots,\theta_k;L)
    = i^{\lt[\frc{k}2\rt]}  \braL\sigma\ketL
    \lt(\prod_{j=1}^k
    \frc{f^{\mu_+}_{\ep_j}(\theta_j;L)}{\braL\sigma\ketL}
    \rt)
    \; \prod_{1\leq i<j\leq k}
    \lt(\tanh\lt(\frc{\theta_j-\theta_i}2\rt)\rt)^{\ep_i\ep_j}~.
\eeq

Similarly, we have
\beq\label{ftfftwistm}
    f^{\Or_-}_{\ep_1,\ldots,\ep_k}(\theta_1,\ldots,\theta_k;L)
    = i^{\lt[\frc{k}2\rt]}  \braL\sigma\ketL
    \lt(\prod_{j=1}^k
    \frc{f^{\mu_-}_{\ep_j}(\theta_j;L)}{\braL\sigma\ketL}\rt)
    \; \prod_{1\leq i<j\leq k}
    \lt(\tanh\lt(\frc{\theta_j-\theta_i}2\rt)\rt)^{\ep_i\ep_j}
\eeq
where $\Or_-$ is $\sigma_-$ if $k$ is even, and $\mu_-$ if $k$ is
odd.

It is easy to check, using (\ref{relftcyl}), that the formulas
above for finite-temperature form factors reproduce the known form
factors on the circle \cite{Bugrij00,Bugrij01,FonsecaZ01}.

Also, it is a simple matter to obtain a Fredholm determinant
representation for the two-point function of twist fields in the R sector.
This is derived in Appendix \ref{appFredholm}.

The zero-temperature form factors of descendant twist fields in
the Majorana theory were analyzed in \cite{CardyM90}, as solutions
to the form factor equations. It was found that the set of form
factors of descendant fields of a given spin $s$ is described by
multiplying form factors of primary twist fields by symmetric
polynomials in $e^{\theta_j},\,j=1,\ldots,k$ (where $\theta_j$ are
the rapidities of the form factor) homogeneous of degree $s$.
Similarly, we expect that finite-temperature form factors of
descendant twist fields are obtained by multiplying those of
primary twist fields by symmetric polynomials. It is easy to check
that the requirements of the Riemann-Hilbert problems of Sections \ref{RHtwist} and
\ref{RHltwist} are still satisfied after such an operation.
However, as explained in the previous section, since rotation is
no longer a symmetry on the cylinder, we cannot identify the spin
of a descendant with the homogeneous degree of symmetric
polynomials. Instead, we must rely on the conjecture in
sub-section \ref{mixingtwist} in order to describe how to combine
appropriate symmetric polynomials. Unfortunately, we have not yet
calculated the mixing matrix for twist fields, hence we cannot
derive more explicit formulas here for descendants.

\sect{Conclusions}

We have derived a method for writing large-distance expansions of
finite-temperature correlation functions. For this purpose, we
introduced the space of operators $\ft$ on which
finite-temperature quantum averages are vacuum expectation values,
and we obtained the expansion of two-point functions by inserting
a resolution of the identity between the two operators. We defined
finite-temperature form factors, which are appropriately
normalized matrix elements on $\ft$ of local operators, and we
described at length their properties. They are related, by
analytical continuation in rapidity space, to form factors in the
quantization scheme on the circle. Any finite-temperature form
factor of a non-interacting field can be seen as the
zero-temperature form factor of a sum of non-interacting fields,
including itself and fields of lower dimension. The coefficients
in this sum are independent of the matrix element evaluated. We
described this phenomenon by constructing the associated mixing
matrix; note that this was overlooked in the initial approach
\cite{LeclairLSS96}. We derived the Riemann-Hilbert problem that
defines the set of all finite-temperature form factors of twist
fields, and calculated finite-temperature form factors of order
and disorder fields.

There are still many open questions:

We have only implicitly described how to obtain large-distance
expansion of correlation functions with more than two operators:
one should insert the resolution of the identity between every
pairs of adjacent operators. The resulting expression involves
matrix elements where both vectors in $\ft$ correspond to excited
states. For non-interacting fields this does not cause any
difficulties, but for twist fields, one must be more careful. We
evaluated such matrix elements of twist fields, as distributions
with terms supported at colliding rapidities (which vanish at zero
temperature) and others supported at separated rapidities. Using
these distributions, all integrals over rapidity variables are
well-defined, and it would be nice to explicitly observe the
agreement of the resulting expression for multi-point correlation
functions with a form factor expansion on the circle. In
particular, from the results presented here, it should be possible
to calculate matrix elements of twist fields in the quantization
on the circle where both vectors correspond to excited states.

We have described how to obtain form factors on the circle from
finite-temperature form factors only for excited states in the NS
sector, where fermion fields have anti-periodic condition around
the cylinder. It would be interesting to describe matrix elements
in a similar way with excited states in the R sector. As we
mentioned, this can be done by defining the inner product on $\ft$
as a trace with a twist operator at, say, position $x=-\infty$ and
having a branch cut on its right.

We have explained how to obtain the mixing matrix associated to
non-interacting fields. In particular, we have obtained an
operator $U$ that describes the action of this matrix on the
non-interacting fields seen as vectors in $\cal L$. The existence
of such an operator is well-known in conformal field theory:
there, the space of operators is isomorphic to the Hilbert space
in radial quantization (which is isomorphic to the Hilbert space
in any quantization on closed lines), and the mixing operator is
described by an operator on the Hilbert space performing a
transformation to the cylinder. It would be interesting to relate
more explicitly the operator $U$ that we wrote with a
transformation to the cylinder.

We obtain finite-temperature form factors of order and disorder
fields. It is important to note that form factors for the right-
and left-operators both specialize to the zero-temperature form
factors only in the region $|{\rm Im}(\theta)|<\pi/2$ of the
complex plane of rapidity variables $\theta$. In connection to
this, observe that the quasi-periodicity relation does not agree
with the one satisfied by zero-temperature form factors. At finite
temperature, this quasi-periodicity relation does not seem to
contain any information about the mutual locality of the twist
field with respect to the fundamental fermion fields. This
semi-locality is included in the thermal poles and zeroes and in
the particular form of the kinematical residue. It would be
interesting to generalize this to other semi-locality index, by
deriving the Riemann-Hilbert problem in the free Dirac theory for
scaling twist fields associated to the $U(1)$ symmetry (in this
connection, see the recent results \cite{Lisovyy05}). But the
thermal poles and zeroes might not be sole consequences of
semi-locality; they might be interpreted as coming from
self-interaction of the field around the cylinder as well, hence
could be present also for fields with zero semi-locality index in
interacting theories.

It could interesting to obtain the non-linear differential
equations describing the two-point function of twist fields from
the Fredholm determinant representation resulting from the
finite-temperature form factor expansion. This would modify some
of the results of \cite{LeclairLSS96}; in particular, it is easy
to see that the equation evolving temperature derivatives would
not hold due to the non-trivial leg factors, and it is not clear
if there is an equation replacing it.

Finally, probably the most important future development from the
ideas presented here is the definition and calculation of
finite-temperature form factors in interacting integrable models,
and the identification of the associated measure of integration in
the resolution of the identity. An important new idea presented
here is the relation between finite-temperature form factors, seen
as matrix elements on the Hilbert space of operators, and form
factors on the circle; in particular, the relation between the
measure of integration and the spectrum on the circle. If, for
fields that are mutually local with respect to fundamental fields,
the finite-temperature form factors equations are the same as the
well-known zero-temperature form factor equations, as was assumed
in previous approaches, then the only non-trivial element is the
measure of integration in the resolution of the identity. This
measure was indeed subject of controversy
\cite{Delfino01,Mussardo01,Castro-AlvaredoF02}. Our ideas give a
guideline for defining this measure of integration. Note, however,
that in view of the works \cite{Smirnov98a,Smirnov98b}, the
rapidity variables might turn out not be the most appropriate ones
for defining finite-temperature form factors.

\bigskip

{\bf Acknowledgments}

I would like to thank J. Cardy, O. Castro Alvaredo, F. Essler and
V. Riva for many insightful discussions and useful comments on the
manuscript, as well as R. Konik for a discussion which lead to an
improvement of the first preprint version of this paper. I
acknowledge support from an EPSRC (UK) post-doctoral fellowship
(grant GR/S91086/01).

\appendix

\sect{Derivation of the relation (\ref{relftcyl})}

\label{appProofftcyl}

A general argument for formula (\ref{relftcyl}) can be obtained
from the relation between traces and expectation values on the
circle, (\ref{ftcorr}) and (\ref{cylcorr}). Recall the expression
of mode operators in terms of fermion operators (\ref{modespsi}).
From this, we can write
\beqa
&&
    {\ }_{+,\dots,+} {\ }^{\hspace{-4mm}\ft}\bra\t\theta_1,\ldots,\t\theta_l|
    \Or^\ft(0,0)|\theta_1,\ldots,\theta_k\ket_{+,\ldots,+}^\ft
    \n
&=&
    \braL a(\t\theta_l) \cdots a(\t\theta_1)
    \Or(0,0) a^\dag(\theta_1) \cdots a^\dag(\theta_k)\ketL
    \; \lt(\prod_{j=1}^l
    \lt(1+e^{-LE_{\t\theta_j}}\rt)\rt)
    \lt(\prod_{j=1}^k \lt(1+e^{-LE_{\theta_j}}\rt)\rt) \n
&=&
    \lt(\frc12\sqrt{\frc{m}\pi}\rt)^k \int dx_1\cdots dx_l\;dx_1\cdots dx_k\;
    e^{-i\sum_{j=1}^l p_{\theta_j}x_j+i\sum_{j=1}^k p_{\theta_j}x_j}\;\times \n
&&
    \times \;\braL \lt(e^{\frc{\t\theta_l}2}\psi(x_l)+ie^{-\frc{\t\theta_l}2}\b\psi(x_l)\rt)
    \cdots
    \lt(e^{\frc{\t\theta_1}2}\psi(x_1)+ie^{-\frc{\theta_1}2}\b\psi(x_1)\rt)
    \Or(0,0) \;\times \n
&&
    \qquad \times\; \lt(e^{\frc{\theta_1}2}\psi(x_1)-ie^{-\frc{\theta_1}2}\b\psi(x_1)\rt)
    \cdots
    \lt(e^{\frc{\theta_k}2}\psi(x_k)-ie^{-\frc{\theta_k}2}\b\psi(x_k)\rt)
    \ketL \;\times \n
&&
    \times \;\lt(\prod_{j=1}^l
    \lt(1+e^{-LE_{\t\theta_j}}\rt)\rt)
    \lt(\prod_{j=1}^k \lt(1+e^{-LE_{\theta_j}}\rt)\rt) \n
&=&
    e^{\frc{i\pi s}2}\lt(\frc12\sqrt{\frc{m}\pi}\rt)^k \int dx_1\cdots dx_l\;dx_1\cdots dx_k\;
    e^{-i\sum_{j=1}^l p_{\theta_j}x_j+i\sum_{j=1}^k p_{\theta_j}x_j}\;\times \n
&&
    \times \;{\ }_L\bra \lt(e^{\frc{\t\theta_l}2+\frc{i\pi}4}\psi_L(x_l)+
        ie^{-\frc{\t\theta_l}2-\frc{i\pi}4}\b\psi_L(x_l)\rt)
    \cdots
    \lt(e^{\frc{\t\theta_1}2+\frc{i\pi}4}\psi_L(x_1)+
        ie^{-\frc{\theta_1}2-\frc{i\pi}4}\b\psi_L(x_1)\rt)
    \Or_L(0,0) \;\times \n
&&
    \qquad \times\; \lt(e^{\frc{\theta_1}2+\frc{i\pi}4}\psi_L(x_1)-
        ie^{-\frc{\theta_1}2-\frc{i\pi}4}\b\psi_L(x_1)\rt)
    \cdots
    \lt(e^{\frc{\theta_k}2+\frc{i\pi}4}\psi_L(x_k)-
        ie^{-\frc{\theta_k}2-\frc{i\pi}4}\b\psi_L(x_k)\rt)
    \ket_L \;\times \n
&&
    \times \;\lt(\prod_{j=1}^l
    \lt(1+e^{-LE_{\t\theta_j}}\rt)\rt)
    \lt(\prod_{j=1}^k \lt(1+e^{-LE_{\theta_j}}\rt)\rt) \n
&=&
    e^{\frc{i\pi s}2}\lt(\frc12\sqrt{\frc{m}\pi}\rt)^k
    {\ }_L\bra\vac|w_-(\t\theta_l)\cdots
    w_-(\t\theta_1) \;\Or_L(0,0) \;w_+(\theta_1)\cdots
    w_+(\theta_k)
    |\vac\ket_L
    \label{B1}
\eeqa
where we define the operators
\beq
    w_\ep(\theta) \equiv \lt(1+e^{- LE_{\theta}}\rt)\;\int dx\; e^{i\ep p_\theta
    x}\lt(e^{\frc{\theta}2+\frc{i\pi}4}\psi_L(0,x)
    -i\ep e^{-\frc{\theta}2-\frc{i\pi}4}\b\psi_L(0,x)\rt)
    ~.
\eeq
We are interested in the analytical continuation
$\theta\mapsto\theta+i\pi/2$ of the matrix element (\ref{B1}) for
all rapidity variables, then in taking the limit where
$\theta\to\alpha_n$ (\ref{defalphan}) of this analytical
continuation. The result of this limit is obtained by taking the
analytical continuation $W_\pm(\theta) = w_\pm(\theta+i\pi/2)$ of
the operators above, then by taking, in all operators
$W_+(\theta)$ on the right of $\Or(0)$, only the negative part of
the integral over $x$, and in all operators $W_-(\theta)$ on the
left of $\Or(0)$, only the positive part of the integral over $x$.
Indeed, the variable $x$ is a time variable in the quantization on
the circle, positive $x$ corresponding to positive time, and the
result of the limit is obtained by looking at the time-ordered
part of the correlation function.

Hence we have
\beqa
W_+(\theta) &\sim&
    \lt(1+e^{-i Lp_{\theta}}\rt) \;\int_{-\infty}^0dx\; e^{-E_\theta
    x}\bigg[ e^{\frc{\theta}2+\frc{i\pi}2}\psi_L(0,x)
    -ie^{-\frc{\theta}2-\frc{i\pi}2}\b\psi_L(0,x)
    \bigg]
     \n
&=&
    \lt(1+e^{-i Lp_{\theta}}\rt)\;\frc{i}{\sqrt{2L}}\sum_{n\in\Z+\frc12}\int_{-\infty}^0 dx\,
        \frc{e^{-E_\theta x}}{\sqrt{\cosh(\alpha_n)}}\;\times\;\n
    && \qquad\qquad\times\;
        \bigg[
            e^{-E_nx}\lt(e^{\frc{\theta+\alpha_n}2} +
                e^{-\frc{\theta+\alpha_n}2}\rt) a_n +
            e^{E_nx}\lt(e^{\frc{\theta+\alpha_n}2} -
                e^{-\frc{\theta+\alpha_n}2}\rt) a_n^\dag
            \bigg] \n
&=&
    \lt(1+e^{-i Lp_{\theta}}\rt)\;\frc{i}{\sqrt{2L}}\sum_{n\in\Z+\frc12}
        \frc{1}{\sqrt{\cosh(\alpha_n)}}\lt[
            -\frc{e^{\frc{\theta+\alpha_n}2} +
                e^{-\frc{\theta+\alpha_n}2}}{E_n+E_\theta} a_n
            +\frc{e^{\frc{\theta+\alpha_n}2} -
                e^{-\frc{\theta+\alpha_n}2}}{E_n-E_\theta} a_n^\dag
            \rt]~. \no
\eeqa
In the last step, we used analytical continuation in the exponents
in order to perform the integral. We can now extract the result of
the limit $\theta\to\alpha_n$:
\beqa
W_+(\theta) &\sim&
    \frc{i}{\sqrt{2L}}
        \frc{2\sinh(\alpha_n)}{\sqrt{\cosh(\alpha_n)}}
        \frc{1+e^{-iLp_{\theta}}}{E_n-E_\theta}\; a_n^\dag
            \n
&\sim&
    \sqrt{2L\cosh(\alpha_n)}\; a_n^\dag~.
\eeqa

On the other hand,
\beqa
W_-(\theta) &\sim&
    \lt(1+e^{-i Lp_{\theta}}\rt) \;\int_0^\infty dx\; e^{E_\theta
    x}\bigg[ e^{\frc{\theta}2+\frc{i\pi}2}\psi_L(0,x)
    +ie^{-\frc{\theta}2-\frc{i\pi}2}\b\psi_L(0,x)
    \bigg]
     \n
&=&
    \lt(1+e^{-i Lp_{\theta}}\rt)\;\frc{i}{\sqrt{2L}}\sum_{n\in\Z+\frc12}\int_0^\infty dx\,
        \frc{e^{E_\theta x}}{\sqrt{\cosh(\alpha_n)}}\;\times\;\n
    && \qquad\qquad\times\;
        \bigg[
            e^{-E_nx}\lt(e^{\frc{\theta+\alpha_n}2} -
                e^{-\frc{\theta+\alpha_n}2}\rt) a_n +
            e^{E_nx}\lt(e^{\frc{\theta+\alpha_n}2} +
                e^{-\frc{\theta+\alpha_n}2}\rt) a_n^\dag
            \bigg] \n
&=&
    \lt(1+e^{-i Lp_{\theta}}\rt)\;\frc{i}{\sqrt{2L}}\sum_{n\in\Z+\frc12}
        \frc{1}{\sqrt{\cosh(\alpha_n)}}\lt[
            \frc{e^{\frc{\theta+\alpha_n}2} -
                e^{-\frc{\theta+\alpha_n}2}}{E_n-E_\theta} a_n
            +\frc{e^{\frc{\theta+\alpha_n}2} +
                e^{-\frc{\theta+\alpha_n}2}}{E_n+E_\theta} a_n^\dag
            \rt] \no
\eeqa
which gives the result of the limit $\theta\to\alpha_n$ to be
\beqa
W_-(\theta) \sim
    \sqrt{2L\cosh(\alpha_n)}\; a_n~.
\eeqa
Hence we recover (\ref{relftcyl}).

\sect{Proof of the equivalence between the finite-temperature form
factor expansion and the form factor expansion on the circle}

\label{appProofexp}

By re-arranging the rapidity variables (along with their
associated charges) in (\ref{ftffexp}), it is possible to bring
the finite-temperature form factor expansion of two-point
functions in the form
\beq\ba{l}
    \braL\Or_1(x,\tau)\Or_2(0,0)\ketL =\z
        \sum_{K=0}^\infty \sum_{k=0}^K
    \int \frc{d\theta_1\cdots
    d\theta_K\;e^{\sum_{j=1}^k (imx\sinh\theta_j-m\tau\cosh\theta_j) -
    \sum_{j=k+1}^K (imx\sinh\theta_j-m\tau\cosh\theta_j)}}{
    k!(K-k)!\;\prod_{j=1}^k\lt(1+e^{-
    mL\cosh\theta_j}\rt)\;\prod_{j=k+1}^{K}\lt(1+e^{
    mL\cosh\theta_j}\rt)}\; \times \z \qquad\qquad\qquad
    \times\;
    f^{\Or_1}_{+,\ldots,+,-,\ldots,-}(\theta_1,\ldots,\theta_k,\theta_{k+1},\ldots,\theta_K;L)
    f^{\Or_2}_{+,\ldots,+,-,\ldots,-}(\theta_K,\ldots,\theta_{k+1},\theta_k,\ldots,\theta_1;L)~.
    \ea
\eeq
If the first field $\Or_1$ is a twist field, we will consider it
to have a branch cut on its right (positive $x$ direction); on the
other hand, if the second field $\Or_2$ is a twist field, we will
consider it to have a branch cut on its left (negative $x$
direction). This ensures that when we obtain the form factor
expansion on the circle, the intermediate states are in the sector
with anti-periodic conditions on the fermion fields (whereas the
vacuum vectors may be in the sector with periodic conditions).
Moreover, we will assume $x>0$; this insures that the operators
are time-ordered in the quantization on the circle.

Consider shifting the contours of integration associated to the
rapidity variables $\theta_{1},\ldots,\theta_k$ towards the
positive imaginary direction, by an amount $i\pi$. The only poles
that contribute are those from the factors
$\prod_{j=1}^k\lt(1+e^{-mL\cosh\theta_j}\rt)$ in the denominator.

Consider the terms arising from taking $N$ poles, with fixed $K$.
They are given by
\beq\ba{l}
    \sum_{k=N}^K
    \int \frc{d\theta_{N+1}\cdots
    d\theta_{K}\;e^{\sum_{j=1}^{N} (-mx\cosh\alpha_{n_j}-im\tau\sinh\alpha_{n_j}) -
    \sum_{j=N+1}^K (imx\sinh\theta_j-m\tau\cosh\theta_j)}}{
    k!(K-k)!\;\prod_{j=N+1}^{K}\lt(1+e^{
    mL\cosh\theta_j}\rt)} \;\times\z\times\;
    (-1)^{k-N}\frc{k!}{N!(k-N)!} \;
    \prod_{j=1}^N\lt(\frc{2\pi}{mL\cosh(\alpha_{n_j})}\rt)
    \; \times \z
    \times\;
    f^{\Or_1}_{+,\ldots,+,-,\ldots,-}\lt(\alpha_{n_1}+\frc{i\pi}2,\ldots,\alpha_{n_N}+\frc{i\pi}2
    ,\theta_{N+1},\ldots,\theta_K;L\rt) \;\times\z\times\;
    f^{\Or_2}_{+,\ldots,+,-,\ldots,-}\lt(\theta_K,\ldots,\theta_{N+1},
    \alpha_{n_N}+\frc{i\pi}2,\ldots,  \alpha_{n_1}+\frc{i\pi}2;L\rt)
    \ea
\eeq
where each element of the set $\{n_j,\,j=1,\ldots,N\}$ is a number
in $\Z+\frc12$. On the third line, there are $N$ positive charges
and $K-N$ negative charges; on the fourth line, there are $K-N$
positive charges and $N$ negative charges. On the second line, the
factor $(-1)^{k-N}$ comes from shifting $k-N$ rapidity variables
by $i\pi$, and taking the imaginary factors from the crossing
relations, (\ref{crossingunint}) and Point 4 in sub-section
\ref{RHtwist}. The sum over $k$ can be done, and vanishes whenever
$N\neq K$. Hence we are left with
\beq\ba{l}
    \frc{e^{\sum_{j=1}^{K} \lt(-mx\cosh\alpha_{n_j}-n_j\frc{2\pi i\tau}L\rt)}}{
    K!} \;
    \prod_{j=1}^K\lt(\frc{2\pi}{mL\cosh(\alpha_{n_j})}\rt)
    \; \times \z
    \times\;
    f^{\Or_1}_{+,\ldots,+}\lt(\alpha_{n_1}+\frc{i\pi}2,\ldots,\alpha_{n_K}+\frc{i\pi}2
    ;L\rt) \;
    f^{\Or_2}_{-,\ldots,-}\lt(\alpha_{n_K}+\frc{i\pi}2,\ldots,\alpha_{n_1}+
    \frc{i\pi}2;L\rt)~.
    \ea
\eeq
When summed over $K$ and over $\{n_j\}$, this reproduces the form
factor expansion on the circle as in (\ref{cylffexp}) if we use
(\ref{relftcyl}), up to a phase factor in accordance with
(\ref{cylcorr}).

\sect{Operators on $\cal H$ corresponding to twist fields}

\label{appTwist}

Let us now turn to the description of the operators associated to
twist fields: the order field $\sigma$ and disorder field $\mu$.

The order and disorder fields can be essentially defined by their
OPE's with the fermion fields. In order to keep the proper OPE's,
it is convenient to remember that with the conformal normalization
$\psi(z)\psi(z')\sim (z-z')^{-1}$, the leading term of the OPE's
are $\psi(z)\sigma(0) \sim
\sqrt{\frc{i}{2z}}\mu(0),\;\psi(z)\mu(0)\sim\sqrt{\frc{-i}{2z}}\sigma(0)$.
Our normalization of the fermions, however, is given by
(\ref{OPEpsipsi}). Hence we have
\beq\label{OPEpsispin}
    \psi(x,\tau)\sigma(0) \sim \frc{i}{2\sqrt{\pi\,(x+i\tau)}}\, \mu(0)~,\quad
    \psi(x,\tau)\mu(0) \sim \frc{1}{2\sqrt{\pi\,(x+i\tau)}} \,\sigma(0)
\eeq
and
\beq\label{OPEpsibarspin}
    \bar\psi(x,\tau)\sigma(0) \sim -\frc{i}{2\sqrt{\pi\,(x-i\tau)}}\, \mu(0)~,\quad
    \bar\psi(x,\tau)\mu(0) \sim \frc{1}{2\sqrt{\pi\,(x-i\tau)}}\,
    \sigma(0)~.
\eeq
By convention, the disorder field has nonzero odd-particle form
factors only, and the order field has nonzero even-particle form
factors only. In particular, $\sigma$ has a nonzero real vacuum
expectation value.

The finite-temperature correlation functions of fermion fields
with insertion of these spin fields are functions on coverings of
the cylinder. In order to define operators on the Hilbert space
$\cal H$ corresponding to the twist fields $\sigma$ and $\mu$, we
must choose Riemann sheets. In fact, since the quantization is on
the line, the cylindrical geometry is implemented by tracing over
$\cal H$, so that we may define the operators corresponding to
twist fields by specifying the Riemann sheets on the plane (that
is, it is sufficient to define them at zero temperature). It will
be convenient to define two operators on $\cal H$ for each twist
field, corresponding to different choices of branch cuts. Inside
vacuum expectation values in $\cal H$, the choice of branch cut
does not matter, but inside traces, it does.

Consider the product of fields $(\psi(x,\tau)\mu(0))_+$ defining
(inside zero-temperature correlation functions) a function on the
plane satisfying the free massive equation of motion everywhere
except at the branch cut $\tau=0,\,x>0$ (and except at the
position of other local fields, if any). We will define the
operator $\mu_+$ by the fact that under the mapping of matrix
elements of operators to correlation functions,
\beqa
    \psi(x,\tau)\mu_+(0) &\mapsto& (\psi(x,\tau)\mu(0))_+\quad
    (\tau>0) \no\\
    \mu_+(0)\psi(x,\tau) &\mapsto& -(\psi(x,\tau)\mu(0))_+\quad
    (\tau<0) \label{eqdebut}
\eeqa
and
\beqa
    \psi(x,\tau)\mu_+(0) &\mapsto& \Big[(\psi(x,\tau')\mu(0))_+\Big]_{\tau':0^+\to\tau}\quad
    (\tau<0) \no\\
    \mu_+(0)\psi(x,\tau) &\mapsto& -\Big[(\psi(x,\tau')\mu(0))_+\Big]_{\tau':0^-\to\tau}\quad
    (\tau>0) ~.
\eeqa
That is, in the second set of maps, the product of operators
gives, inside vacuum expectation values, correlation functions
that are continued through the branch cut of
$(\psi(x,\tau')\mu(0))_+$ if $x>0$.

Similarly, consider the product of fields $(\psi(x,\tau)\mu(0))_-$
defining a function on the plane that satisfies the free massive
equation of motion everywhere except at the branch cut
$\tau=0,\,x<0$ (and except at the position of other local fields,
if any); this function coincides with the function defined by
$(\psi(x,\tau)\mu(0))_+$ when $\tau>0$. We will define the
operator $\mu_-$ by the fact that under the mapping of matrix
elements of operators to correlation functions,
\beqa
    \psi(x,\tau)\mu_-(0) &\mapsto& (\psi(x,\tau)\mu(0))_-\quad
    (\tau>0) \no\\
    \mu_-(0)\psi(x,\tau) &\mapsto& -(\psi(x,\tau)\mu(0))_-\quad
    (\tau<0)
\eeqa
and
\beqa
    \psi(x,\tau)\mu_-(0) &\mapsto& \Big[(\psi(x,\tau')\mu(0))_-\Big]_{\tau'=0^+\to\tau}\quad
    (\tau<0) \no\\
    \mu_-(0)\psi(x,\tau) &\mapsto& -\Big[(\psi(x,\tau')\mu(0))_-\Big]_{\tau'=0^-\to\tau}\quad
    (\tau>0)~.
\eeqa

We make similar definitions for the order field, without the minus
sign that represents the odd statistics of $\mu_\pm$ with fermion
operators. We have:
\beqa
    \psi(x,\tau)\sigma_+(0) &\mapsto& (\psi(x,\tau)\sigma(0))_+\quad
    (\tau>0) \no\\
    \sigma_+(0)\psi(x,\tau) &\mapsto& (\psi(x,\tau)\sigma(0))_+\quad
    (\tau<0)
\eeqa
and
\beqa
    \psi(x,\tau)\sigma_+(0) &\mapsto&
        \Big[(\psi(x,\tau')\sigma(0))_+\Big]_{\tau'=0^+\to\tau}\quad
    (\tau<0) \no\\
    \sigma_+(0)\psi(x,\tau) &\mapsto&
        \Big[(\psi(x,\tau')\sigma(0))_+\Big]_{\tau'=0^-\to\tau}\quad
    (\tau>0)
\eeqa
as well as
\beqa
    \psi(x,\tau)\sigma_-(0) &\mapsto& (\psi(x,\tau)\sigma(0))_-\quad
    (\tau>0) \no\\
    \sigma_-(0)\psi(x,\tau) &\mapsto& (\psi(x,\tau)\sigma(0))_-\quad
    (\tau<0)
\eeqa
and
\beqa
    \psi(x,\tau)\sigma_-(0) &\mapsto&
        \Big[(\psi(x,\tau')\sigma(0))_-\Big]_{\tau'=0^+\to\tau}\quad
    (\tau<0) \no\\
    \sigma_-(0)\psi(x,\tau) &\mapsto&
        \Big[(\psi(x,\tau')\sigma(0))_-\Big]_{\tau'=0^-\to\tau}\quad
    (\tau>0)~. \label{eqfin}
\eeqa

With these definitions and with the OPE's (\ref{OPEpsispin}) and
(\ref{OPEpsibarspin}), it is possible to check that the fields
$\sigma_\pm$ are Hermitian, that $\mu_+$ is Hermitian and that
$\mu_-$ is anti-Hermitian on $\cal H$:
\beq
    \sigma_\pm^\dag = \sigma_\pm~,\quad \mu_\pm^\dag = \pm\mu_\pm~.
\eeq

Note that relations (\ref{eqdebut} -- \ref{eqfin}) also hold for
operators corresponding to descendants of the order and disorder
fields.

\sect{Normalization of the one-particle finite-temperature form
factors}

\label{appnorm}

The normalization can be obtained by considering the OPE's
$\psi(x,\tau)\mu_\pm(0)$. Consider first $\mu_-(0)$. We have
\beq\label{Tpsimu-}
    \braL {\cal T} \psi(x,\tau)\mu_-(0)\ketL \sim \frc1{2\sqrt{\pi (x+i\tau)}}
    \braL\sigma\ketL
\eeq
where $\cal T$ means time ordering (the latest operator being
placed on the left) and where the square root is taken on its
principal branch. We will take $\tau>0,\,x>0$ in (\ref{Tpsimu-});
then we can repeat the calculation of Appendix \ref{appProofexp}.
Since only the one-particle finite temperature form factors
contribute, the results of Appendix \ref{appProofexp} give, for
the left-hand side,
\beq
    \sum_{n\in\Z+\frc12} e^{-mx\cosh\alpha_{n}-\frc{2i \pi n \tau}L}\;
    \frc{2\pi}{mL\cosh(\alpha_{n})}\;
    f^{\psi}_{+}\lt(\alpha_{n}+\frc{i\pi}2;L\rt) \;
    f^{\mu_-}_{-}\lt(\alpha_{n}+\frc{i\pi}2;L\rt)~.
\eeq
The leading $x+i\tau\to0^+$ behavior is obtained by looking at the
leading $|n|\to\infty$ behavior of the summand. We have
\[
    f_-^{\mu_-}\lt(\alpha_{n}+\frc{i\pi}2;L\rt) \sim
    e^{-\frc{i\pi}4} C(L)~.
\]
From (\ref{ftfffermions}), the main contribution comes from the
region $n\to\infty$ only, and we obtain for the leading behavior
of the left-hand side of (\ref{Tpsimu-}),
\beq
    \frc{C(L)}{\sqrt{L}}\sum_{n\in\Z+\frc12,\,n>0} e^{-\frc{2\pi n(x+i\tau)}L}\;
    \frc{1}{\sqrt{n}}\;
     \sim \sqrt{\frc{1}{2(x+i\tau)}} \; C(L)~.
\eeq
Hence, we find (\ref{normC}).

It is instructive to repeat the calculation for $\mu_+(0)$ and to
verify that $C(L)$ is indeed as given above. We have
\beq\label{Tpsimu+}
    \braL {\cal T} \psi(x,\tau)\mu_+(0)\ketL \sim \frc1{2\sqrt{\pi (x+i\tau)}}
    \braL\sigma\ketL
\eeq
where the square root is on a branch that coincides with the
principal branch for $\tau>0$ but that has a cut at
$\tau=0,\,x>0$. In order to apply the results of Appendix
\ref{appProofexp}, we need take $\tau<0,\,x<0$, and we have
\beq\label{Tpsimu+2}
    \braL \mu_+(0)\psi(x,\tau)\ketL \sim i\frc1{2\sqrt{-\pi (x+i\tau)}}
    \braL\sigma\ketL
\eeq
where now the square root is on its principal branch. This has
expansion
\beq
    \sum_{n\in\Z+\frc12} e^{mx\cosh\alpha_{n}+\frc{2i \pi n \tau}L}\;
    \frc{2\pi}{mL\cosh(\alpha_{n})}\;
    f^{\mu_+}_{+}\lt(\alpha_{n}+\frc{i\pi}2;L\rt) \;
    f^{\psi}_{-}\lt(\alpha_{n}+\frc{i\pi}2;L\rt)~.
\eeq
With $f_+^{\mu_+}(\theta;L)\sim e^{\frc{i\pi}4} C(L)$ and using
(\ref{ftfffermions}), we have, for the leading $x+i\tau\to 0^-$
behavior of the left-hand side,
\beq
    \frc{i\, C(L)}L \sum_{n\in\Z+\frc12,\,n>0} e^{\frc{2 \pi n (x+i\tau)}L}\;
    \frc{1}{\sqrt{n}} \sim i \sqrt{\frc{1}{-2(x+i\tau)}} \; C(L)
\eeq
which agrees with (\ref{Tpsimu+2}) if $C(L)$ is given by
(\ref{normC}).

\sect{Fredholm determinant representations for two-point functions
at finite temperature}

\label{appFredholm}

The results of Appendix \ref{appProofexp} essentially show that
\beq\ba{c}
    \braL \sigma_+(x,\tau)\sigma_-(0,0)\ketL = {\
    }_L\bra\vac|\sigma_L(-\tau,x) \sigma_L(0,0)|\vac\ket_L \\
    \braL \mu_+(x,\tau)\mu_-(0,0)\ketL = {\
    }_L\bra\vac|\mu_L(-\tau,x) \mu_L(0,0)|\vac\ket_L
    \ea
\eeq
where on the right-hand side, the vacuum expectation values are on
the circle in the Ramond sector. In a similar fashion, an analysis
of the finite-temperature form factor expansion shows that
\beq\ba{c}
    \braL \sigma_-(x,\tau)\sigma_+(0,0)\ketL =
    \braL \sigma_+(-x,-\tau)\sigma_-(0,0)\ketL \\
    \braL \mu_-(x,\tau)\mu_+(0,0)\ketL =
    - \braL \mu_+(-x,-\tau)\mu_-(0,0)\ketL~.
    \ea
\eeq
Note that the second equation is in agreement with the fermionic
statistic of the operators $\mu_\pm$.

Using the finite-temperature form factors (\ref{ftfftwistp}) and
(\ref{ftfftwistm}) for twist fields, we have then the following
large distance expansions of two-point functions in the R sector:
\beq\label{R1}\ba{l}
    \braL\sigma_+(x,\tau)\sigma_-(0,0)\ketL = \z
        \sum_{k=0\atop k\ {\rm even}}^\infty \sum_{\ep_1,\ldots,\ep_k=\pm}
    \int \frc{d\theta_1\cdots
    d\theta_k\;e^{\sum_{j=1}^k\ep_j (imx\sinh\theta_j-m\tau\cosh\theta_j)}}{
    k!\;\prod_{j=1}^k\lt(1+e^{-\ep_j
    mL\cosh\theta_j}\rt)}\;
    i^k \prod_{j=1}^k (f^{\mu_+}_{\ep_j}(\theta_j;L))^2
    \prod_{1\leq i<j\leq k}
    \tanh\lt(\frc{\theta_j-\theta_i}2\rt)^{2\ep_i\ep_j}
    \ea
\eeq
and
\beq\label{R2}\ba{l}
    \braL\mu_+(x,\tau)\mu_-(0,0)\ketL =\z
        -\sum_{k=0\atop k\ {\rm odd}}^\infty \sum_{\ep_1,\ldots,\ep_k=\pm}
    \int \frc{d\theta_1\cdots
    d\theta_k\;e^{\sum_{j=1}^k\ep_j (imx\sinh\theta_j-m\tau\cosh\theta_j)}}{
    k!\;\prod_{j=1}^k\lt(1+e^{-\ep_j
    mL\cosh\theta_j}\rt)}\;
    i^k \prod_{j=1}^k (f^{\mu_+}_{\ep_j}(\theta_j;L))^2
    \prod_{1\leq i<j\leq k}
    \tanh\lt(\frc{\theta_j-\theta_i}2\rt)^{2\ep_i\ep_j}.
    \ea
\eeq
Following \cite{BabelonB92,LeclairLSS96}, Fredholm determinant
representations can now easily be obtained from the formulas
\beq\label{f1}
    {\rm det}_{i,j} \lt\{ \frc{u_i-u_j}{u_i+u_j} \rt\} =
    \lt\{\ba{ll} \prod_{1\leq i<j\leq k}
    \lt(\frc{u_i-u_j}{u_i+u_j}\rt)^2
    & k \ {\rm even} \\
    0 & k \ {\rm odd}
     \ea
    \rt.
\eeq
and
\beq\label{f2}
    {\rm det}_{i,j} \lt\{\frc1{u_i+u_j}\rt\} =
    \frc1{2^ku_1\ldots u_k} \prod_{1\leq i<j\leq k}
    \lt(\frc{u_i-u_j}{u_i+u_j}\rt)^2~.
\eeq
Formula (\ref{f1}) gives
\beq
    \braL\sigma_+(x,\tau)\sigma_-(0,0)\ketL
    = {\rm det}({\bf 1} + {\bf K})
\eeq
where ${\bf K}$ is an integral operator with an additional index
structure, defined by its action $({\bf K} f)_\ep(\theta) =
\sum_{\ep'=\pm} \int_{-\infty}^{\infty} d\theta'
K_{\ep,\ep'}(\theta,\theta')f_{\ep'}(\theta')$ and its kernel
\beq
    K_{\ep,\ep'}(\theta,\theta') =
    i (f_\ep^{\mu_+}(\theta;L))^2
    \tanh\lt(\frc{\theta'-\theta}2\rt)^{\ep\ep'}\;
    \frc{e^{\ep(imx\sinh\theta-m\tau\cosh\theta)}}{1+e^{-\ep
    mL\cosh\theta}}~.
\eeq
Finally, in order to obtain two-point functions of disorder
fields, we must consider the linear combinations $\sigma\pm \mu$.
Formula (\ref{f2}) gives
\beq
    \braL(\sigma_+(x,\tau)+\eta\mu_+(x,\tau))(\sigma_-(0,0)+\eta\mu_-(0,0))\ketL
    = {\rm det}({\bf 1} + {\bf J}^{(\eta)})
\eeq
with $\eta=\pm$ and by definition $({\bf J}^{(\eta)}f)_\ep(u) =
\sum_{\ep'=\pm} \int_{0}^{\infty} du'
J^{(\eta)}_{\ep,\ep'}(u,u')f_{\ep'}(u')$ where the kernel is given
by
\beq
    J^{(\eta)}_{\ep,\ep'}(u,u') =
    -2\eta\, i (f_\ep^{\mu_+}(\ln(u);L))^2
    \frc1{\ep u + \ep' u'}\;
    \frc{e^{\frc{\ep}2(i m x(u-u^{-1})-m\tau(u+u^{-1}))}}{1+e^{-\frc{\ep
    m L}2(u+u^{-1})}}~.
\eeq

\end{document}